\documentclass[11pt]{article}
\usepackage{color,amsfonts,amsmath,amssymb,mathrsfs,verbatim,epsf}

\def\inn{\in}

\def\TM{\mbox{\it TM}}

\def\rrr{{\mathbb{R}}}
\def\ccc{{\mathbb{C}}}

\def\ooo{{\mathbb{O}}}

\def\zzz{{\mathbb{Z}}}
\def\qqq{{\mathbb{Q}}}

\def\b1{\mbox{\boldmath $1$}}

\def\v{\mbox{\boldmath $v$}}

\def\bv{\mbox{\boldmath $v$}}
\def\bvh{\hat{\bv}}
\def\hG{\hat{G}}

\def\bx{\mbox{\boldmath $x$}}

\def\lvh{L(\hat{\v})=1}

\def\lvt{L(\v_{27})=1}
\def\lvfs{L(\v_{56})=1}

\def\htwc{\mbox{h}_2\ccc}

\def\htho{\mbox{h}_3\ooo}

\def\sltc{\mbox{SL}(2,\ccc)}

\def\sltho{\mbox{SL}(3,\ooo)}

\def\uo{\mbox{U}(1)}

\def\sutw{\mbox{SU}(2)}

\def\suth{\mbox{SU}(3)}

\def\ee{\mbox{E}_8}

\def\ese{\mbox{E}_7}

\def\esi{\mbox{E}_6}

\def\GN{G_{\! N}}

\def\gpath{ }
\def\setb{\setlength{\baselineskip}{0.625\baselineskip}}

\begin{document} 

%% take out double spacing (28/11/15)
{\setlength{\baselineskip}{0.625\baselineskip}

\begin{center}

{\LARGE{\bf  The Structure of Matter in Spacetime from}} \\
 \vspace{5pt}
 {\LARGE{\bf  the Substructure of Time}} \\

   \vspace{6pt}

 \mbox {{\Large David J. Jackson}}\footnote{email: david.jackson.th@gmail.com}  \\

  \vspace{4pt}
 
%\today
 { \large February 22, 2018}

  \vspace{4pt}

{\bf  Abstract}

\vspace{-11pt} 
 
\end{center}

 The nature of the change in perspective that accompanies the proposal of a unified physical theory %%@
deriving from the single dimension of time is elaborated. On expressing a temporal interval in a %%@
multi-dimensional form, via a direct arithmetic decomposition, both the geometric structure of %%@
4-dimensional spacetime and the physical structure of matter in spacetime can be derived from the %%@
substructure of time. While reviewing this construction, here we emphasise how the new conceptual picture %%@
differs from the more typical viewpoint in theoretical physics of accounting for the properties of matter %%@
by first postulating entities on top of a given spacetime background or by geometrically augmenting %%@
4-dimensional spacetime itself. With reference to historical and philosophical sources we argue that the %%@
proposed perspective, centred on the possible arithmetic forms of time, provides an account for how the %%@
mathematical structures of the theory can relate directly to the physical structures of the empirical %%@
world.

{

\vspace{-19pt}
\tableofcontents
\vspace{-12pt}

% for footnote space
%\vspace{0.9cm}
%\vspace{15pt}
}

% putthis at bottom to take out double line spaceing for all
%\par}% \linespread{1.0} for main text (28/11/15)
%match '{\setlength{\baselineskip}{0.625\baselineskip}' above

%\pagebreak

\section{Introduction}
\label{pers1}

   Building upon the initial proposal of Kaluza~\cite{Kaluza} and Klein~\cite{Klein} %%@
there are many ways in which a theory can be constructed over 4-dimensional spacetime %%@
by utilising extra spatial dimensions. The options include for example the number of %%@
extra dimensions, the properties of the overall geometric structure and the manner in %%@
which physics in the familiar four dimensions is extracted. This is reflected in the %%@
extensive literature on the subject as reviewed for example in %%@
\cite{Rizzo,Shif,Cheng,Liu} and the references therein.
 However, despite the many approaches, the geometric structures of extra spatial %%@
dimensions do not readily lead to known empirical properties of the Standard Model of %%@
particle physics without some difficulty in contriving such features (see for example %%@
\cite{Witt,Jitt}).  
\begin{comment}
 with one of the main goals in recent decades being to explain how structures of the %%@
Standard Model of particle physics can derive from the geometric structures of the %%@
extra dimensions  (see for example \cite{Witt,Jitt}). 
\end{comment} 
  In particular the origin of the distinctive symmetry patterns of Standard Model %%@
particle multiplets remains a puzzle.

   In this paper we describe how a physical theory can be derived from the seemingly %%@
counter-intuitive starting point of a single temporal dimension only. This can be %%@
achieved through exploiting the basic arithmetical structure of the real line, as %%@
representing progression in time. In contrast with postulating an initial %%@
higher-dimensional structure from which physics in 4-dimensional spacetime is %%@
extracted, building the theory up from the one dimension of time alone provides a %%@
well-defined and much more restricted basis for a theory. 
 One striking feature of this new approach~\cite{Unifi,Novel,KKone,TimeE} is the %%@
ability to directly account for significant elements of Standard Model structure on %%@
adopting this change of perspective.

  In the following section we describe the contrast in the conceptual basis in more %%@
detail and explain how it is possible to construct a theory from this simple idea %%@
based on time. In section~\ref{pers3} the explicit development of the theory and %%@
explanatory power that can be achieved is summarised, based on the technical details
 described in~\cite{Unifi,Novel,KKone,TimeE}. The main focus of this paper is the %%@
underlying change in perspective regarding the relationship between space, time and %%@
matter that accompanies the mathematical development and physical successes of the %%@
theory. The core arguments for this change in emphasis from extra spatial dimensions %%@
to the one dimension of time are elaborated in section~\ref{pers4}. The underlying %%@
motivation is further considered in section~\ref{pers5} in comparison with that of %%@
other physical theories and unification schemes, before reiterating the role of this %%@
paper in the context of~\cite{Unifi,Novel,KKone,TimeE} in the conclusions.

%\pagebreak

\section{Space, Time, Matter; a Choice in Perspective}
\label{pers2}

  From an objective point of view events in the physical world 
  take place at locations in space and time and
   are typically ascribed to the properties of matter, postulated with the aim of %%@
accounting for empirical phenomena observed on all scales as sketched in %%@
figure~\ref{arena}(c). The 4-dimensional spacetime manifold $M_4$, as the arena for %%@
such events at locations $x \in M_4$, is pictured by itself in figure~\ref{arena}(a).
 This spacetime manifold possesses a metric tensor $g(x)$, as signified by the %%@
rectangular frames in figure~\ref{arena},  defining a \mbox{light cone} structure on %%@
$M_4$. With components $g_{\mu\nu}(x)$, for general coordinate indices \mbox{$\mu,\nu %%@
= 0,1,2,3$}, the metric $g(x)$ may be considered either globally or only locally %%@
equivalent to the Minkowski metric $\eta = \mbox{diag}(+1,-1,-1,-1)$, depending upon %%@
the theoretical framework.
 A physical theory might then be constructed by introducing further entities upon %%@
$M_4$ as the background `stage', or by geometrically augmenting $M_4$ itself with %%@
extra spatial dimensions with particular properties. Alternatively we might first ask %%@
where the mathematical structure for $(M_4,g)$ itself arises from, that is, what %%@
supports the stage itself?

\begin{figure}[htbp]  
\centering
\epsfxsize=14.4cm
\leavevmode
\epsffile[0 0 2435 755]{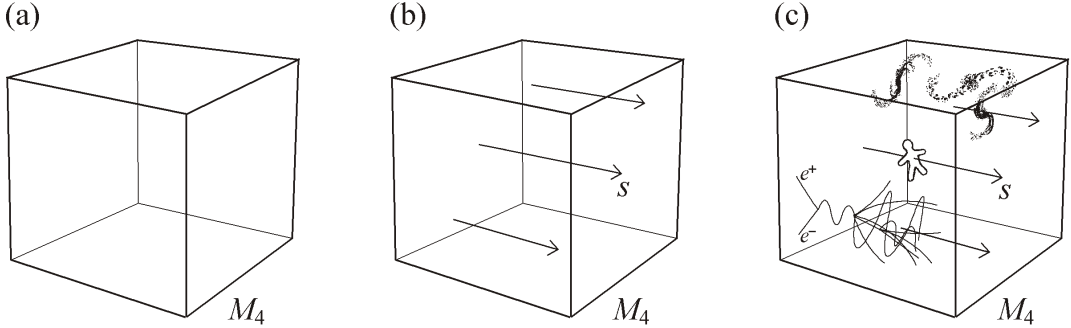}
\caption{\setb  (a) The empty stage of spacetime $(M_4,g)$, with one spatial dimension %%@
suppressed in the sketch, (b) contains a universal flow of time $s$ which, through the %%@
change in perspective that we describe in this paper, can be considered as the %%@
mathematical progenitor of 4-dimensional spacetime itself as well as (c) the matter %%@
content on all scales from particle physics to cosmology.}
\label{arena}
\end{figure}

  As an irreducible feature of this spacetime arena events on $M_4$ are infused with %%@
the flow of time, in that any `clock' at any location and on any physical trajectory %%@
within the light cone structure will `tick' as a record of the passage of time, which %%@
can be more generally parametrised by a real number $s \in \rrr$ as depicted in %%@
figure~\ref{arena}(b). From a subjective point of view we also note that this passage %%@
of time accompanies any observer, as represented in the centre of %%@
figure~\ref{arena}(c), and any observations that we can make in the universe.

   The metric structure of spacetime, as incorporated into general relativity %%@
consistent with the equivalence principle, implies that at any spacetime location %%@
local inertial coordinates $(x^0,x^1,x^2,x^3)$ may be identified such that an
 infinitesimal proper time interval $\delta s$ between two proximate events (such as %%@
two infinitesimally separated `ticks of a clock') can be expressed, within a local %%@
Lorentz transformation, as:
\begin{equation}
 \label{propint}
 (\delta s)^2 \; = \; (\delta x^0)^2 \, - \, (\delta x^1)^2
              \, - \, (\delta x^2)^2 \, - \, (\delta x^3)^2  
\end{equation}  

  Hence a one-dimensional flow of time $s\in \rrr$ can be conceived of as completely %%@
filling the 4-dimensional spacetime volume $M_4$ in figure~\ref{arena}(b) while being %%@
everywhere locally expressed in the form of equation~\ref{propint}. This universal %%@
temporal flow is analagous to `cosmic time' in models for the evolution of the %%@
universe. However we initially consider figure~\ref{arena}(b) to represent the flow of %%@
time through a flat, empty spacetime; that is there are global coordinate systems in %%@
which the metric can be globally equated with the Minkowski metric $g(x) = \eta = %%@
\mbox{diag}(+1,-1,-1,-1)$ implicit in equation~\ref{propint} and,
 with no other structure identified on $M_4$, this manifold describes the arena of %%@
special relativity.
 With respect to the metric the local Lorentz invariant $\delta s$ can be interpreted %%@
as a timelike interval  \textit{constructed from} four local coordinate intervals in %%@
equation~\ref{arena} relating 3-dimensional space $(x^1,x^2,x^3)$ and time $(x^0)$  in %%@
a unified `spacetime' structure.

 The alternative perspective that we are proposing is to consider the extended %%@
4-dimensional spacetime manifold $M_4$ itself in figure~\ref{arena}(b) to be a %%@
possible \textit{manifestation of} the one-dimensional temporal flow $s\in \rrr$. That %%@
is, on taking $s \in \rrr$ as the \textit{primary} entity in equation~\ref{propint} %%@
this quadratic expression \textit{for} an infinitesimal interval of time implies that %%@
time \textit{itself} contains an elementary substructure and symmetry that underlies %%@
the construction of the extended spacetime manifold $M_4$ pictured in %%@
figure~\ref{arena}(b). While the arithmetic substructure of time in %%@
equation~\ref{propint}  expresses the geometric structure of an infinitesimal local %%@
inertial coordinate frame the Minkowski metric implicit in that equation (which is %%@
equivalent to equation~\ref{propeta} below) is exhibited throughout an implied %%@
extended manifold $M_4 \equiv \rrr^4$ through the translation symmetry over the full %%@
range of  $(x^0,x^1,x^2,x^3) \in \rrr^4$ for that equation (as described in detail %%@
in~\cite{TimeE} opening of subsection~4.1 and figure~1 there). Hence the globally flat %%@
manifold $M_4$ in figure~\ref{arena}(b) can be interpreted as a representation %%@
\textit{of} the flow of time.

  Since the `equals' sign in equation~\ref{propint} makes no statement about the %%@
priority of the left or right-hand side in this expression, we are free to make this %%@
alternative interpretation with $\delta s \in \rrr$ identically containing the %%@
arithmetic substructure of the basis \textit{for} an extended 4-dimensional spacetime. %%@
In this sense `spacetime', practically as the name might suggest, can be interpreted %%@
as an augmentation of the flow of time itself. 
We can obtain equation~\ref{propint} by initially writing $\delta s = \delta x^0 \in %%@
\rrr$ and $(\delta s)^2 = (\delta x^0)^2$ and then opening up this trivial expression %%@
for time directly through the arithmetic modification $(\delta s)^2 = (\delta %%@
x^0)^2-\vert\delta \bx\vert^2$,
  incorporating a 3-vector 
   $\delta\bx = (\delta x^1,\delta x^2,\delta x^3) \in \rrr^3$ of further real %%@
numbers,
on introducing Lorentz transformations algebraically as a `symmetry of time' leaving %%@
$\delta s$ invariant. This allows the flow of time to be simultaneously manifested in %%@
a geometric spacetime form as depicted in figure~\ref{arena}(b), with \textit{time} %%@
$\delta s$ hence incorporating \textit{space} through this direct arithmetic %%@
composition. 
 In particular we can consider equation~\ref{propint}  to represent an infinitesimal %%@
interval of time $\delta s \neq 0$ associated with an observer and a local inertial %%@
reference frame \textit{constructed} through this 4-dimensional expression %%@
\textit{for} time with Lorentz and translational symmetries.

  Hence here the Lorentz symmetry on the spacetime manifold $M_4$ enters the theory as %%@
a symmetry of time that preserves the ordered causal property of time within the light %%@
cone structure on $M_4$. With equation~\ref{propint} written as
  $(\delta s)^2 = (\delta x^0)^2-\vert\delta \bx\vert^2$ the Lorentz transformations %%@
leave $\delta s$ invariant for \textit{any} timelike trajectory with $\delta s \neq %%@
0$, with Lorentz boosts changing the ratio $\vert \delta \bx \vert \! : \! \delta x^0$ %%@
and hence the `speed' through space. However any physical entity that propagates %%@
within this causal structure on $M_4$ but along a null-trajectory, on the light cone %%@
itself with $\delta s = 0$, will have the \textit{same} 
   $\vert \delta \bx \vert \! : \! \delta x^0$ ratio in any Lorentz frame and hence %%@
the same speed. This will be the case for example for an electromagnetic or a %%@
gravitational wave, and in particular the constancy of the speed of light for any %%@
reference frame hence follows as a consequence.

 This contrasts with the basis of special relativity founded upon the postulate of 
the `principle of relativity' -- that the equations of physics, and in particular of %%@
electrodynamics as well as mechanics, should take the same form in any inertial %%@
reference frame and hence transform in a covariant manner between such frames in %%@
uniform relative motion -- together with the further \textit{postulate} that the speed %%@
of light in empty space should be the same in any inertial frame, independent of the %%@
motion of the light 
 source~(\cite{Einsr}, \cite{Pais} chapter~7).
 A departure from Galilean relativity is then required to accommodate the second  %%@
postulate, with the Lorentz transformations between inertial reference frames in %%@
uniform relative motion shown to derive from these first principles in maintaining
$(\delta s)^2 = (\delta x^0)^2-\vert\delta \bx\vert^2 = 0$ 
  for light propagation.
 With Maxwell's equations for electrodynamics then expressible in Lorentz covariant %%@
form the apparent conflict between the theories of Newton and Maxwell was resolved.

  While for special relativity, founded on the constancy of the speed of light with %%@
$\delta s = 0$, space ($\bx$) and time ($x^0$) coordinates enter on an equal footing, %%@
mutually mixing under Lorentz transformations in a unified spacetime structure, here %%@
the founding motivation places the priority on a
 causally ordered
 fundamental flow of time ($s$)
 with $\delta s \neq 0$ associated with the perspective of a local observer,
  with space intrinsically embedded within this flow via equation~\ref{propint}.
  For the full theory the local observer, as depicted in figure~\ref{arena}(c), may or %%@
may not be associated with a local inertial reference frame while, 
  similarly as for general relativity, 
 the extended Minkowski base space of special relativity, and of %%@
figure~\ref{arena}(b), arises in the flat spacetime limit.

   To develop the full theory, building upon equation~\ref{propint},
  we can then consider what form might be taken by a further, more general, %%@
augmentation as an arithmetic expression for the flow of time. Given that we are %%@
generalising from equation~\ref{propint} this can also be interpreted as broadening %%@
the notion of a local inertial coordinate frame of general relativity. The form of the %%@
local proper time interval $\delta s$ 
 can be generalised from the 4-dimensional Lorentzian form of equation~\ref{propint} %%@
(for which $a,b = 0,1,2,3$ here can be interpreted as indices for a local inertial %%@
coordinate frame):
\begin{eqnarray}
    (\delta s)^2 & = & \eta_{ab }\delta x^a \delta x^b  \qquad\qquad 
	    \mbox{with} \quad\qquad
               \eta = \mbox{diag}(+1,-1,-1,-1)    \label{propeta}  \\
    \mbox{to} \quad 
	(\delta s)^p & = & \alpha_{abc\ldots}\delta x^a \delta x^b \delta x^c \ldots
	\qquad \mbox{with each} \qquad \alpha_{abc\ldots} \in \{-1,0,1\}  \label{propgen}
\end{eqnarray} 
  as a homogeneous $p^{\mathrm{th}}$-order polynomial in $n$ dimensions. With %%@
component labelling of $0, \ldots, (n-1)$ or $1,\ldots,  n$ being a matter of %%@
convention, in equation~\ref{propgen} the sum is taken over each index  $a,b,c\ldots = %%@
1, \ldots ,n$  for the real intervals
   $\{\delta x^1, \ldots , \delta x^n\} \in \rrr^n$ 
   \mbox{(\cite{TimeE} equations~40--42)}.

  With emphasis upon the temporal interval $\delta s$ on the left-hand side of these %%@
expressions, the generalisation in equation~\ref{propgen} is \textit{not} required to %%@
be of a quadratic spacetime form with a higher-dimensional Lorentz symmetry. In %%@
particular, cubic and higher-order homogeneous polynomial forms are also permitted. 
  While equation~\ref{propgen} is hence more general than the approach of adding extra %%@
spatial dimensions, the development of the theory is more constrained in that we %%@
\textit{add nothing else} beyond the structure and symmetries of explicit forms for %%@
equation~\ref{propgen} in directly deducing the basic structure of the physical %%@
theory. These observations mark the main contrast with the range of models based upon %%@
extra spatial dimensions alluded to in the opening of section~\ref{pers1}.
The conceptual and philosophical aspects of this shift in perspective will be %%@
considered further in section~\ref{pers4} as the central theme of this paper.

 While equation~\ref{propgen} describes a possible higher-dimensional %%@
\textit{mathematical} form of time, we still \textit{see} the physical world in the %%@
form  of the arena of the 4-dimensional manifold $M_4$ of figure~\ref{arena} with the %%@
local metric structure of equation~\ref{propeta}. With the form of 3-dimensional space %%@
incorporated within 4-dimensional spacetime, the necessary identification of the %%@
extended structure $M_4$ with a local Lorentz symmetry
 \textit{breaks} the full symmetry of the higher-dimensional form in %%@
equation~\ref{propgen} (see for example~\cite{TimeE}~subsection~4.1). The residual %%@
`extra dimensions' of equation~\ref{propgen}, over and above 4-dimensional spacetime, %%@
can then be interpreted as underlying the structure of matter in spacetime,
 in principle accounting for empirical phenomena such as sketched in %%@
figure~\ref{arena}(c). That is, under this change in perspective, time is not %%@
considered a benign independent variable flowing through the world but rather, via the %%@
substructure of equation~\ref{propgen}, as the antecedent source of both spacetime and %%@
the matter it contains. Rather than being a mere spectator the flow of time is %%@
simultaneously manifested as the background stage together with the theatrical scenery   %%@
and cast of characters upon it.

 The question then concerns the degree to which the substructure of time, as expressed %%@
through equation~\ref{propgen}, might account for the structure of matter.
 Such augmentations from equation~\ref{propeta} allow the incorporation of the %%@
non-flat external spacetime of general relativity in association with further physical %%@
features on $M_4$,
  with for example the manifestly covariant form of Maxwell's source-free equations %%@
proposed to arise as reviewed for~(\cite{Unifi} equations~5.29 and 5.30) and implying %%@
the constant speed of light propagation.
 More generally
  the theory might then be tested on all scales of the universe, from the microscopic %%@
world to the expanses of cosmological structure, as recorded by the observer, all %%@
collectively progressing in time as depicted in figure~\ref{arena}(c).  
 In particular a correspondence between the substructure of time and
  the properties of elementary particles as observed in high energy physics %%@
experiments may be sought.

  In considering possible symmetries of higher-dimensional forms of time it is %%@
convenient to express equation~\ref{propgen} in terms of the generally finite %%@
components, defined by
$v^a = \frac{dx^a}{ds} = \frac{\delta x^a}{\delta s} 
   {\big{\vert}}_{\mbox {\tiny $\delta s \! \to \! 0$}}$ for $a=1\ldots n$,
  of an $n$-dimensional vector $\bv_n \inn \rrr^n$ 
    by rearranging that equation, on dividing both sides by $(\delta s)^p$ and taking %%@
$\delta s \to 0$, to obtain
 the homogeneous polynomial form denoted:
\begin{equation}
  L(\bv_n)  \; := \;    
    \alpha_{abc\ldots}\,  v^a\, v^b\, v^c \ldots  \; = \; 1
    \label{lvo}
\end{equation}
 The origin of this expression, as the central equation of the theory, is also %%@
described for (\cite{Unifi} equation~2.9, \cite{Novel} equation~11, \cite{KKone} %%@
equation~4 and \cite{TimeE} equation~43).
  
  While $\bv_n \in \rrr^n$ in equation~\ref{lvo} represents the  $n$ components of the %%@
full $n$-dimensional form of time considered, the subcomponents
  $\bv_4 = (v^0,v^1,v^2,v^3) \in \TM_4$ represent the projection out of the full set %%@
in  $\rrr^n$ onto the tangent space of $M_4$, as an intrinsic feature of the symmetry %%@
breaking structure. (In general the components of $\bv_4\in \TM_4$ need not be in %%@
one-to-one correspondence with a conventional list of subcomponents of $\bv_n \in %%@
\rrr^n$, see for example the discussion of~\cite{TimeE} equation~54).  In the %%@
following section we briefly review the more detailed structure of this symmetry %%@
breaking for explicit mathematical forms for equation~\ref{lvo} and summarise the %%@
resulting physical structures on $M_4$ that are derived, before returning in %%@
section~\ref{pers4} to elaborate upon the conceptual perspective adopted for this %%@
theory.

%\pagebreak

\section{Explicit Substructure of Time; Elements of Physics}
\label{pers3}

 In seeking explicit higher-dimensional forms for equation~\ref{lvo} with a high %%@
degree of symmetry natural extensions from the quadratic Lorentzian form (rewriting  %%@
equations~\ref{propint} and \ref{propeta} in the form of equation~\ref{lvo}):
\begin{equation}
 \label{lvfr}
  L(\bv_4)\, = \, (v^0)^2 - (v^1)^2 - (v^2)^2 - (v^3)^2 \, = \,
   \eta_{ab }v^a  v^b \, = \, 1
\end{equation}
   lead to an $\esi$ symmetry of the 27-dimensional \textit{cubic} form:
\begin{equation}
 \label{lvts}
  L(\bv_{27}) = \det(\bv_{27}) = 1 \qquad \mbox{with} \qquad 
                     \bv_{27} \in \htho
\end{equation}
 where elements of $\htho$, the exceptional Jordan algebra, are Hermitian $3\times 3$ %%@
matrices over the octonions $\ooo$  (\cite{Unifi} chapters~6 and 8, \cite{Novel} %%@
sections~4 and 5, \cite{KKone} subsection~5.2, \cite{TimeE} subsection~4.2, adopting %%@
mathematical structures described for example in \cite{Man4,Wang2}). 
 In general we then have $L(\bv_4) = h^2 \neq 1$, with $h\inn \rrr$, as a substructure %%@
embedded within $\lvt$ (as noted for~\cite{Unifi} equations~5.46 and 13.1), with a %%@
similar generalisation applying for subsequent augmentations.
In turn the symmetry and structure of equation~\ref{lvts} can be further embedded %%@
within an $\ese$ symmetry acting upon the 56-dimensional \textit{quartic} form:
\begin{equation}
 \label{lvfs}
  L(\bv_{56}) = q(\bv_{56}) = 1 \qquad \mbox{with} \qquad 
                     \bv_{56} \in F(\htho)
\end{equation}
  with the $4^{\mathrm{th}}$-order form $q$ defined on the Freudenthal triple system
   $F(\htho)$ (\cite{Unifi} section~9.2, \cite{Novel} section~6, \cite{KKone} %%@
subsection~5.2, \cite{TimeE} subsection~4.3, based upon mathematical structures %%@
described for example in \cite{Borst} section~9, \cite{Rios}). 

In the case of equation~\ref{lvfs} the symmetry breaking projection of $\bv_4 \in %%@
\TM_4$ over the external spacetime $M_4$ out of $\bv_{56} \in F(\htho)$ leads to %%@
transformation properties of the reduced 56-dimensional representation space under the %%@
resulting external Lorentz and residual internal gauge symmetry that describe `matter %%@
fields' bearing a close resemblance to structures of the Standard Model as summarised %%@
in table~\ref{esebreak}.
  In particular Lorentz spinor structures are identified as well as internal $\suth_c$ %%@
colour singlets and triplets with the appropriate fractional charges under an internal %%@
$\uo_Q$ associated with electromagnetism.  
\def\rai{+0.3ex}
\begin{table}[htbp]
\centering
\begin{tabular}{|r|ccccc|c|}
 \hline
  \raisebox{-0.5ex}{$\mathbf{56}$} \!\!\!\!\!\!\!
   {\mbox{\raisebox{+0.0ex}{\LARGE{$\diagdown$}}}} \!\!\!\!\!\!
      \mbox{\raisebox{+0.7ex}{$\ese \! \supset\!\!$}} 
	    & \raisebox{\rai}{Lorentz} & \raisebox{\rai}{$\!\!\!\times\!\!\!$}
	    & \raisebox{\rai}{$\suth_c$}   & \raisebox{\rai}{$\!\!\!\times\!\!\!$} 
		& \raisebox{\rai}{$\uo_Q$} & \raisebox{\rai}{Matter} \\
 \hline
  $\mathbf{8} \qquad$  & Dirac & & $\mathbf{1}$ & & $1$ & $e$-lepton  \\
 $\mathbf{24} \qquad$  & Dirac & & $\mathbf{3}$ & & $\frac{1}{3}$ & $d$-quark  \\
  $\mathbf{4} \qquad$  & \underline{vector} & &
                   $\mathbf{1}$ & & $0$ & $\nu_L$-lepton  \\
 $\mathbf{12} \qquad$  & \underline{scalar} & & 
                   $\mathbf{3}$ & & $\frac{2}{3}$ & $u$-quark  \\
  $\mathbf{4} \qquad$  & \underline{vector} & & 
                   $\mathbf{1}$ & & $0$ & Higgs  \\
  $\mathbf{4} \qquad$  & scalar & & $\mathbf{1}$ & & $0$ & Dark  \\
 \hline   
  \end{tabular}
  \caption{\setb The structures deriving from the breaking of the $\ese$ symmetry of %%@
\mbox{$\lvfs$} in equation~\ref{lvfs} with their interpretation as `matter fields' %%@
listed in the final column.
 In identifying this correspondence with a generation of Standard Model leptons and %%@
quarks there is very little redundancy -- with the four additional scalar singlet %%@
components providing `dark sector' candidates. (See the discussion of 
 \protect\cite{TimeE} \mbox{figure~4 for details).}}
\label{esebreak}
\end{table} 

 In addition elements of $\sutw_L \times \uo_Y$ electroweak theory can be identified %%@
in the symmetry breaking structure from the $\esi$ level of equation~\ref{lvts} %%@
(\cite{Unifi} section~8.3), with a direct left-right asymmetry emerging at the $\ese$ %%@
level of equation~\ref{lvfs} and table~\ref{esebreak} (\cite{Unifi} equation~9.46, %%@
\cite{Novel} equation~66, \cite{TimeE} subsection~4.3). Unlike the case for the other %%@
lepton and quark states 
 the neutral components associated with the neutrino can only be accommodated in %%@
either the left \textit{or} right-handed sector of the fragmented $\mathbf{56}$ %%@
components and is hence denoted $\nu_L$ in table~\ref{esebreak}, being complementary %%@
to the projected $\bv_4 \in \TM_4$ components (as discussed in~\cite{TimeE} shortly %%@
before figure~4).
 The Standard Model Higgs is conjectured to be intimately related to the external %%@
4-vector $\bv_4 \in \TM_4$ itself and the symmetry breaking projection over $M_4$  %%@
(for reasons explained in \cite{TimeE} shortly after figure~4).

The need to address the discrepant external Lorentz symmetry transformation %%@
properties, underlined in table~\ref{esebreak}, and to account for a full three %%@
generations of leptons and quarks incorporating a full electroweak theory and %%@
associated Higgs phenomena, leads to the prediction  of a significant role for the %%@
largest exceptional Lie group $\ee$ as the full symmetry of time, as emphasised in %%@
\cite{TimeE}.

  While the possibility of connections between the above exceptional Lie groups and %%@
structures of the Standard Model is well known (see for example \cite{Gur1}, %%@
\cite{Gur7} and \cite{Lisi} for examples involving $\esi$, $\ese$ and $\ee$ %%@
respectively, each however with a symmetry breaking pattern that differs from that %%@
considered for the present theory), here we begin with a well-defined conceptual %%@
starting point which motivates equation~\ref{lvo} as the pivotal expression through %%@
which these symmetries and their breaking patterns might be realised in the physical %%@
world. Via equation~\ref{lvo} for $n>4$ the flow of time is channelled through the %%@
external dimensions of the spacetime manifold $M_4$ and the residual dimensions of an %%@
internal space, the absolute distinction of which implies an absolute breaking of the %%@
full symmetry, denoted $\hat{G}$ (taken as $\hat{G}=\esi$, $\ese$ or $\ee$ for %%@
example), which is reduced to the direct product:     
\begin{equation} 
  \label{dirprod}
   \mbox{Lorentz} \times G \, \subset \, \hG
\end{equation}

  Here the external Lorentz symmetry may be expressed via its double cover $\sltc$, as %%@
for the $\ese$ case (more explicitly described in \cite{TimeE} figure~4) for which the %%@
external tangent space is embedded in the higher-dimensional spaces as $\TM_4 \equiv %%@
\htwc \subset \htho \subset F(\htho)$, hence accommodating the Dirac spinor %%@
representations listed in 
table~\ref{esebreak}. The residual  group factor
  $G$ in equation~\ref{dirprod} is the internal symmetry, where we have $G=\suth_c %%@
\times \uo_Q \subset \ese$
  in  table~\ref{esebreak} for this 56-dimensional form of time. 

   Out of the full $n$-dimensional space implicitly underlying the $n$-dimensional %%@
form of time $\lvh$, where $\hat{\bv} = \bv_n$ for the largest $n$ considered, only a %%@
4-dimensional subset, exhibiting the quadratic form in equation~\ref{lvfr}, is %%@
perceived as an extended geometrical manifold, namely $M_4$ as depicted for example in %%@
figure~\ref{arena}(b). This structure incorporates the `spatialisation of time', as %%@
described by 3-dimensional spacelike hypersurfaces embedded within $M_4$, which on %%@
account of this spatial property can hence be directly diagrammatically visualised, %%@
albeit with one spatial dimension suppressed in the sketch of $M_4$ in %%@
figure~\ref{arena}. This describes the arena within which the structures of the %%@
physical world within which we are immersed arise, with the properties of matter %%@
determined by the structure and symmetries of the additional components over %%@
4-dimensional spacetime as summarised in table~\ref{esebreak} for the $\ese$ level. %%@
These latter internal algebraic structures do not have a literal geometric %%@
interpretation and hence, unlike the external dimensions, cannot be described by a %%@
direct visual representation other than as a distribution of `matter' in spacetime, as %%@
sketched in figure~\ref{arena}(c).

  Within this symmetry breaking structure, on utilising the translational symmetry of  %%@
equations~\ref{propint} and \ref{lvfr} in $(x^0,x^1,x^2,x^3) \in \rrr^4 \equiv M_4$ %%@
for the corresponding subcomponents of the higher-dimensional form $\lvh$, a globally %%@
flat spacetime can again be generated as depicted in figure~\ref{arena}(b) with %%@
Einstein tensor $G^{\mu\nu}(x) = 0$
(the Einstein tensor is defined in general in terms of components of the Riemann %%@
curvature tensor, see for example~\cite{KKone} subsection~3.1). However,
 more generally now a relation between the external curvature, associated with the %%@
Lorentz symmetry, and an internal curvature, associated with the residual gauge %%@
symmetry $G$ of equation~\ref{dirprod}, analogous to that constructed in non-Abelian %%@
Kaluza-Klein theories, can be identified  with $G^{\mu\nu}(x) \neq 0$
  (\cite{Unifi} equation~5.20, \cite{KKone} equation~93)
   from the geometric structure over $M_4$ resulting from the symmetry breaking.
  Further,
 allowing for both variations in and exchanges between the various field components  %%@
for the augmented form of time, a global solution with generally non-zero external %%@
curvature can be considered as expressed by
 (\cite{Unifi}~equations~5.32 and 15.1, \cite{TimeE} equation~85):
\begin{equation}
 \label{gfromavt}
  G^{\mu\nu} \: = \: f(A,\bvh)   \: =: \:  -\kappa T^{\mu\nu}   
\end{equation} 
    which also \textit{defines} the energy-momentum tensor $T^{\mu\nu}(x)$, with %%@
$\kappa$ a normalisation constant.
 In these cases the Minkowski metric $\eta$ of equation~\ref{lvfr} strictly only %%@
applies on the manifold $M_4$ for local inertial frames as for general relativity.
 In equation~\ref{gfromavt} the tensor function $f$ denotes a  composition of fields %%@
arising from the symmetry breaking over $M_4$, with a more specific structure for this %%@
function to be determined by the constraints of the theory and the full form of time %%@
employed.
  The component $A(x)$ represents gauge fields associated with the internal symmetry %%@
$G$, such as
 $\suth_c \times \uo_Q \subset \ese$ in table~\ref{esebreak}, while $\bvh(x)$ %%@
represents the fragmented components of the multi-dimensional form of time, as also %%@
listed in  table~\ref{esebreak} for the case of the broken $\ese$ symmetry action on
 the 56-dimensional representation underlying the form $\lvfs$.

 Both the multiplet structure of Standard Model particle states, described by a matter %%@
content such as listed in table~\ref{esebreak}, and their quantum mechanical %%@
properties, associated with a local degeneracy in spacetime solutions for %%@
equation~\ref{gfromavt}, are proposed to arise through this breaking of the full %%@
symmetry $\hG$ of the full form $\lvh$ over the 4-dimensional substructure composing %%@
the spacetime arena $M_4$,  
with the Lorentz component of the reduced symmetry in equation~\ref{dirprod}
  acting on $\TM_4$. 
  Through the indeterminacy of field solutions underlying the local external spacetime %%@
geometry $G^{\mu\nu}(x) \neq 0$ in equation~\ref{gfromavt} the
  apparent quantum particle transitions associated with the physical properties of %%@
matter both arise and are accessible to observation, for example through the event %%@
processes of high energy physics experiments such as that sketched in %%@
figure~\ref{arena}(c).   
   In general the physical properties of matter and the interactions through which we %%@
observe it are associated with the energy-momentum tensor $T^{\mu\nu}(x)$, 
  as defined in equation~\ref{gfromavt}, the form of which hence derives from   
  the possible compositions of the solution $G^{\mu\nu}  =  f(A,\bvh)$ for spacetime %%@
itself.

    Historically the difficulties in combining general relativity and quantum theory %%@
in a single consistent mathematical framework has proved practically intractable.
 However, consistent with the existence of an underlying unifying conceptual %%@
structure, the empirical world itself seamlessly exhibits the properties of both of %%@
these principal pillars of modern physics, albeit as empirically verified in typically %%@
complementary observational environments.
 An understanding of how the two theoretical frameworks of general relativity and %%@
quantum theory might be extracted from a single unifying theory is likely to involve a %%@
more complete understanding of the nature of the connection between spacetime geometry %%@
and the properties of matter, whether or not the latter has its origins in a structure %%@
of `extra dimensions'. This connection is proposed to arise here through %%@
equation~\ref{gfromavt} with the principles of general relativity largely preserved, %%@
and with the Einstein field equation:
\begin{equation}
 \label{Eineq}
    G^{\mu\nu} = -\kappa T^{\mu\nu}
\end{equation}
  contained within that expression. On the other hand the machinery of quantum theory %%@
is conjectured to arise from the indeterminacy implied in the field composition %%@
$f(A,\bvh)$ in equation~\ref{gfromavt} under the idealisation and approximation of a %%@
flat spacetime limit.

  Standard postulates for the quantum theory of matter are formulated against a flat %%@
space or spacetime arena. However according to general relativity the presence of %%@
matter is associated with a non-flat spacetime through the Einstein field %%@
equation~\ref{Eineq}. Here we interpret this association to apply on all scales and %%@
hence standard quantum theory based on the assumption of a flat spacetime environment  %%@
\textit{can} only represent an approximation.
  In attempting to combine gravitation and quantum theory the approximation of %%@
Newton's theory of gravity is not generally employed for the former, and here we %%@
suggest that applying the apparent approximations of quantum theory, in particular to %%@
`quantise' the gravitational field which describes deviations from a flat spacetime, %%@
is similarly insufficient for a complete theory. Instead we propose that it is gravity %%@
itself, through the degeneracy of spacetime solutions for equation~\ref{gfromavt}, %%@
that provides  the fundamental mechanism for the quantisation of all non-gravitational %%@
fields. Both classical general relativity and standard quantum theory can in principle %%@
consistently emerge as limiting cases from the unifying conceptual picture of this %%@
theory (\cite{Unifi} chapter~11, \cite{KKone} subsection~5.3, 
  \cite{TimeE} section~6).

   While Newtonian gravity has been superseded by Einstein's theory we do not possess %%@
a framework to supplant quantum theory, and hence here we are proposing such a %%@
possibility through equation~\ref{gfromavt}.  
The mismatch between the higher symmetry $\hG$ of the full temporal form $\lvh$ and %%@
the lower local Lorentz symmetry of spacetime $M_4$, breaking the former to the %%@
product of equation~\ref{dirprod} as the full multi-dimensional form of time is %%@
necessarily filtered through the external 4-dimensional frame of all observations %%@
depicted in figure~\ref{arena}, 
 is central to the resulting quantum nature of matter. 
  This universal symmetry reducing structure, proposed as the origin of the quantum %%@
properties of matter, is closely analogous to the further symmetry reducing  %%@
conditions imposed by the measurement apparatus through which particular quantum %%@
phenomena are typically observed in the laboratory, as noted with reference to the %%@
Zeeman effect in (\cite{Unifi} section~11.4).

  As noted above the identification of an extended 4-dimensional spacetime manifold %%@
$M_4$, with the local Minkowski metric structure of equation~\ref{lvfr}, breaks the %%@
full symmetry $\hG$ of the full multi-dimensional \textit{mathematical} form of time %%@
$\lvh$ from equation~\ref{lvo} \textit{absolutely} down to the direct product subgroup %%@
of equation~\ref{dirprod} as the surviving symmetry for \textit{physical} structures %%@
in the spacetime arena $M_4$.
 Hence the \textit{full} unifying symmetry $\hG$ of the theory is hidden and %%@
\textit{not} accessible for empirical phenomena.
 In particular we note that the structure of a direct product of the external Lorentz %%@
and internal symmetry $G$ in equation~\ref{dirprod} for the physics is compatible with %%@
the Coleman-Mandula theorem~\cite{ColMan} for non-trivial particle scattering in the %%@
relativistic quantum theory limit.

  On the gravitational side in general relativity there is an ambiguity in the meaning %%@
of the Einstein equation~\ref{Eineq} in terms of the relative priority of the geometry %%@
$G^{\mu\nu}(x)$ or the energy-momentum $T^{\mu\nu}(x)$ on either side of that %%@
expression. The original view of Einstein was influenced by Mach's principle -- as %%@
expressed through the complete determination of the metric $g_{\mu\nu}(x)$, underlying %%@
the Einstein tensor $G^{\mu\nu}(x)$,
  by the mass distribution of physical bodies, or more generally $T^{\mu\nu}(x)$ %%@
(\cite{Pais}~sections~15(e~and~f), \cite{Tor} section~6.2). 
 With `distant masses and their motions\ldots regarded as the seat of the causes' %%@
(\cite{Eingr} section~2), and while in turn the gravitational field guides the course %%@
of material processes, general relativity was not proposed as a `theory of matter' %%@
(\cite{Eingr} section~18).
  A similar view generally prevails today as expressed through the 
   popular interpretation of the Einstein equation  that `spacetime is warped by %%@
matter', that is `$G^{\mu\nu} \leftarrow T^{\mu\nu}$'.
 This interpretation also has its origins in Newton's theory for which mass can be %%@
considered as the source of gravitation and which can be identified in the appropriate %%@
limit of Einstein's theory, indeed with the constant in equation~\ref{Eineq} expressed %%@
as \mbox{$\kappa = 8\pi \GN/c^4$} where $\GN$ is Newton's gravitational constant.

 Here, on the contrary, in equation~\ref{gfromavt} we place the priority on possible %%@
solutions for the spacetime geometry $G^{\mu\nu} = f(A,\bvh)$  with the %%@
energy-momentum $T^{\mu\nu}$ being essentially \textit{defined} through this %%@
expression, that is `$G^{\mu\nu} \rightarrow T^{\mu\nu}$'. 
(A similar interpretation of the Einstein equation~\ref{Eineq} is found for example in %%@
\cite{Schr} chapter~XI, although still with assumptions needed for an effective matter %%@
Lagrangian to shape the specific empirical properties of matter).
 We might then also turn around the connection with Mach's principle, with geometric %%@
solutions $G^{\mu\nu} = f(A,\bvh)$ being of primary concern and with the inertial %%@
properties of matter $T^{\mu\nu}$ \textit{derived} as a consequence of the nature of %%@
these solutions for equation~\ref{gfromavt}. 
 Indeed the geometric Bianchi identity for $G^{\mu\nu}(x)$ itself effectively implies %%@
both energy-momentum conservation (as discussed in \cite{Unifi} opening of %%@
section~5.2) and the geodesic flow of matter (as reviewed for \cite{Unifi} %%@
equation~5.36). 
While shaped by gravity the inertial properties of matter are made apparent through an %%@
interplay with the other forces of nature such as electromagnetism, since the %%@
non-gravitational forces can act \textit{within} a local inertial reference frame.
 These more general empirical structures of matter derive from the restrictive forms %%@
for \mbox{$f(A,\bvh) =:  -\kappa T^{\mu\nu}$}  in the possible solutions for %%@
equation~\ref{gfromavt}, as permitted within the constraints of the theory (as %%@
described for \cite{Unifi} equation~11.29), in principle \textit{avoiding} the need to %%@
postulate a matter Lagrangian entirely.
 The explicit development of this theory has led to the specific structures listed in %%@
table~\ref{esebreak}, incorporating features such as fractional charges and a %%@
left-right asymmetry that resemble the Standard Model. Hence through %%@
equations~\ref{lvo} and \ref{gfromavt} here we do have a `theory of matter', and one %%@
which more generally  
 applies for matter on all scales as sketched in figure~\ref{arena}(c), as will be %%@
discussed further in the following section.
This has been achieved by stepping back from spacetime and matter and founding the %%@
theory ultimately on the  flow of time alone.

  The ambiguity in the meaning of the `equals' sign in equation~\ref{Eineq} is %%@
analogous to that for equation~\ref{propint}. In the latter case here we also give %%@
priority to the left-hand side, that is to $\delta s$ in expressing an interval of %%@
time. In the following section we further discuss the motivation, meaning and %%@
consequences of adopting this latter choice of perspective in placing the emphasis on %%@
the underlying one-dimensional flow of time.

%\pagebreak

\section{`Gestalt Shift' in Viewpoint on the Universe}
\label{pers4}

 From our familiar world of 3-dimensional space it is very difficult, and perhaps %%@
impossible, to imagine what geometric structures would \textit{look like} for %%@
inhabitants of a world with four or more extended dimensions of space. However it is %%@
also very  hard, if not impossible, to conceive in our mind's eye of how things would %%@
look from the \textit{intrinsic} perspective of beings in a world of only two spatial %%@
dimensions, or even one alone, without a sense of that world being embedded or %%@
constrained within our native 3-dimensional spatial framework. On the other hand we %%@
can readily conceive of progression through the one dimension of time, essentially the %%@
vehicle by which all of our thought processes are conveyed. From this subjective point %%@
of view the experience of a coherent, self-contained, one-dimensional progression in %%@
time, accompanying all of our observations, can also proceed in the absence of any %%@
impression of spatial dimensions, as for example in focussing upon the sound of a %%@
piece of music in a dark place.  
  It is this one-dimensional continuum of time, represented by a progression of values %%@
on the real line $s\in \rrr$, that provides the original connection with the left-hand %%@
side of equations~\ref{propint}, \ref{propeta} and \ref{propgen}, and hence directly %%@
with the arithmetic substructure on the right-hand side of these expressions that is %%@
subsequently exploited by the theory described in this paper. 

 Objectively we typically treat three dimensions of space 
along with a fourth dimension parametrised by a timelike component 
  as given together in the spacetime arena $M_4$, as pictured in %%@
figure~\ref{arena}(b), which \textit{accommodates} the passage of time $s\in \rrr$
 subject to equation~\ref{propint}. It is perhaps from the subjective point of view %%@
that the shift in perspective we are considering regarding figure~\ref{arena}(b) is %%@
more readily made, with all of our thoughts and observations taking place through the %%@
passage of time, through which in turn a spacetime arena can be \textit{constructed} %%@
via the substructure in equation~\ref{propint}, as reviewed in section~\ref{pers2}.

 From the philosophical point of view the `gapless continuum' of time is in a sense %%@
necessary to hold a thought together as `a thought' or as part of a continuous %%@
coherent `train of thoughts'. We might then begin by considering
 Descartes' position of sceptical doubt in reducing certainties about the world to a %%@
minimal `I think therefore I am' (\cite{Desc} part IV), adapted rather  to the %%@
proposal  that `I think therefore time exists', as a generic observation for the %%@
possibility of there being thoughts.  
   In contrast to Descartes' philosophical argument for constructing a theory of %%@
knowledge of the external empirical world
   from the `I am', as expounded in his book `Meditations on First Philosophy' of %%@
1642, here we construct a full physical theory of the world from the minimalist %%@
starting point of the existence of time. 
  Analysis of the
  arithmetic substructure of the real line representing time allows the construction %%@
of such a theory which might be tested against empirical phenomena observed in the %%@
physical world.
At an elementary level this is possible since the arithmetic properties implicit in a %%@
real interval include operations of multiplication and division, opening up a richer %%@
substructure than addition and subtraction alone, as utilised for example in %%@
equation~\ref{propint}.

 As a mathematician Descartes was also one of the first to note a correspondence %%@
between these basic arithmetic operations ($+,-,\times,\div$, together with the %%@
extraction of roots) and geometric constructions (\cite{Desc} appendix~3 `La %%@
G\'eom\'etrie', see also \cite{Smith} part~III on Descartes). This work contributed to %%@
the invention of coordinate geometry as also developed independently in the 1630s by %%@
Pierre de Fermat (\cite{Smith} part III on Fermat). 
 Through this connection algebra can be applied to model and solve geometric problems %%@
using mathematical methods also known as analytic geometry.
Here we consider that the quadratic composition in the real components 
 $(\delta x^1,\delta x^2,\delta x^3)$ for the possible arithmetic substructure of time %%@
in equation~\ref{propint} can provide the basis for the geometrical form of a %%@
\textit{physical} 3-dimensional 
 Euclidean space \textit{itself}, arbitrarily extended through the 
  $(x^1,x^2,x^3) \inn \rrr^3 $ translation symmetry.

In addition to the necessity of time as an \textit{a priori} form for all thoughts and %%@
experiences, the \textit{a priori} necessity of space as an arena to frame the %%@
physical objects we perceive was emphasised by the philosopher Kant in the %%@
1780s~\cite{Kant}. 
Here we are proposing that the arithmetic form of time itself in %%@
equation~\ref{propint}  provides our \textit{a priori} predisposition to perceive the %%@
world in space as well as time with the appropriate underlying mathematical framework %%@
as something to `get hold of'.
  That is, the quadratic structure of this multi-dimensional temporal form
     provides the equivalent mathematical representation underlying the geometric %%@
construction of the Pythagorean theorem and the formulation of Euclid's postulates 
 pertaining to a continuous, indefinitely extended, homogeneous and isotropic, metric %%@
space.
 Hence  the flow of time \textit{carries with it}, via equation~\ref{propint}, an %%@
arithmetic substructure that \textit{can} be simultaneously apprehended `externally' %%@
in the form of an extended geometric spatial arena, providing the necessary background %%@
for our perceptions of the physical world.

 The more general properties of an $n$-dimensional manifold space with a metric %%@
geometry were elaborated by Riemann in 1854 (\cite{Riem}, see also \cite{Smith} part %%@
III on Riemann), with the case of 
  physical space described as a triply  extended magnitude  upon which the square of %%@
an indefinitely small line element $\delta s$ can be expressed locally as 
$(\delta s)^2 = \sum_{i=1}^3(\delta x^i)^2$. Hence the Pythagorean theorem still holds %%@
in the limit of small scale geometric structure.  
  In order to analyse the mathematical structure of global metric relations Riemann %%@
introduced methods of tensor analysis, including the Riemann curvature tensor as a %%@
measure of the deviation from flatness of the manifold.
  Riemann noted that the assumption of a global Euclidean spatial geometry with 
   zero curvature is only a hypothesis and not a certainty when, in the %%@
mid-$19^{\mathrm{th}}$ century, extrapolating beyond the bounds of observation both on %%@
the very large and very small scale. He also speculated upon the possible implications %%@
this may have for physics and raised the question of the source of metric relations. %%@
In particular Riemann conjectured that components of the curvature of space can have %%@
arbitrary values for the smaller scales, provided the total curvature of the region is %%@
close to zero, and 
 continues (towards the end of \cite{Riem} and \cite{Smith} part III on Riemann, here %%@
quoting from the latter):
\begin{quotation}
   Even greater complications may arise in case the line element is not representable, %%@
as has been premised, by the square root of a differential expression of the second %%@
degree.
\end{quotation}

  By 1916 Einstein had proposed the energy-momentum of matter as the source of %%@
geometric curvature through the Einstein equation~\ref{Eineq} and established his %%@
theory of gravity~\cite{Eingr}, as we described towards the end of the previous %%@
section.
  To achieve this Einstein employed a 4-dimensional manifold with the local Minkowski %%@
metric structure of equation~\ref{propeta}, which we can write as
 $(\delta s)^2 = (\delta x^0)^2-\sum_{i=1}^3(\delta x^i)^2$, to incorporate the %%@
equivalence principle with special relativity holding in the limit of small local %%@
inertial reference frames.
 Hence, while the metric of Riemannian geometry is positive definite, both  special %%@
and general relativity adopt the generalisation of a pseudo-Riemannian or Lorentzian 
  manifold with a non-degenerate but indefinite 
  metric for 4-dimensional `spacetime'. This smooth symmetric metric $g$ describes a %%@
light cone structure on the spacetime manifold $(M_4,g)$ as discussed for %%@
figure~\ref{arena}.
 Although the `line element' is now identified with the `proper time interval'
  $\delta s$ 
  a quadratic structure for the line element  is maintained in the theory of %%@
relativity in equations~\ref{propint} and \ref{propeta}.

  In the present theory we are generalising further, echoing the above quote from %%@
Riemann, and are led to higher-order homogeneous forms for the `line element' on %%@
interpreting equations~\ref{propint} and \ref{propeta} as a `form of time' consistent %%@
with the general expression in equation~\ref{propgen}, which is not restricted to %%@
homogeneous polynomials of the second degree.
 That is, since we are taking the perspective of placing the emphasis on the left-hand %%@
side of these expressions, with `time' taking priority over `space', the %%@
generalisation from $(\delta s)^2$ to $(\delta s)^p$, with $p>2$, is a natural one.
 Here we simply analyse the possible basic arithmetic forms of time which, as well as %%@
incorporating the spatial form in equation~\ref{propeta}, exhibit the more general %%@
structures and symmetries of equation~\ref{propgen}, or equivalently %%@
equation~\ref{lvo}.

 The higher-order polynomial forms still contain quadratic substructures that underpin %%@
our perception of spatial structures.
The full symmetry $\hG$ of the full form of time $\lvh$ is then broken
 through the \textit{a priori} requirement of perceiving the world not only through %%@
time but also in the form of space, as elucidated by Kant,  described now by Euclidean %%@
geometry \textit{locally} and still to within a good approximation on the macroscopic %%@
scale. This symmetry breaking in identifying the 4-dimensional spacetime background %%@
$M_4$ is described for
 equation~\ref{dirprod} 
 and leads directly to the microscopic structure of matter via %%@
equation~\ref{gfromavt}.
  That is,  the properties of the `extra dimensions' in equation~\ref{lvo}, over and %%@
above those needed to construct the 4-dimensional spacetime manifold, are manifested %%@
as the physical structures of matter subject to laws of physics that might be deduced %%@
from the constraints of the theory.
  Pursuing this idea  for the 56-dimensional form of time $\lvfs$ of %%@
equation~\ref{lvfs}, and analysing the breaking pattern of the full $\ese$ symmetry, %%@
has led to non-trivial success with a series of empirical Standard Model properties %%@
identified as summarised for table~\ref{esebreak}.

  Since the flow of time through the general multi-dimensional form of %%@
equations~\ref{lvo} is perceived as motion in space through the extended 4-dimensional %%@
geometric substructure $M_4$, as depicted in figure~\ref{arena}(c), and since `matter' %%@
is in part defined as that which `occupies space', the practical interpretation of %%@
observations in the world naturally leads to conceptions of material substance and its %%@
interactions. 
  For any theory of matter
such interactions are proposed to account for both the apparent properties of matter %%@
and our ability to make observations of them.

In one of the first books on general relativity Weyl suggests the general definition: %%@
`In the wide sense, in which we now use the word, matter is that of which we take %%@
cognisance directly through our senses' (\cite{Weyl2} section~25). More precisely Weyl %%@
also notes that we can `assign the term matter to that real thing, which is %%@
represented by the energy-momentum tensor' (\cite{Weyl2} section~25), that is %%@
$T^{\mu\nu}$ on the right-hand side of the Einstein equation~\ref{Eineq}.
  In the historical context of the early 1920s Weyl describes how 
 this practical definition can incorporate a theory in which the basic elements are %%@
fields. `Matter' might then be considered `an offspring of the field' with the 
atomic properties of matter, including electron phenomena, associated with 
  `energy-knots'  of localised extreme values propagating 
	 in the electromagnetic field  (\cite{Weyl2} section~25).

	  In the present theory both the classical and quantum properties of matter are %%@
incorporated in $T^{\mu\nu}$ as \textit{defined} in equation~\ref{gfromavt} through %%@
the degeneracy of field solutions under $f(A,\bvh)$ for the external spacetime %%@
geometry $G^{\mu\nu}$.
 The resulting properties of matter then include the elementary particle phenomena %%@
observed in high energy physics experiments with the range of possible particle %%@
interactions shaped by the internal substructure of the flow of time itself as %%@
described for example for table~\ref{esebreak}.

  Ordinary macroscopic matter is not then to be considered as `built out of' %%@
elementary particles, rather matter on all scales is a direct manifestation of %%@
mathematical relations deriving from the multi-dimensional forms of the underlying %%@
flow of time, with the latter essentially perceived as a `flow of matter' through %%@
spacetime within which we are immersed.
  The conception of a microscopic material particle substratum derives from the %%@
process of breaking up `matter' -- so to account for the macroscopic properties of %%@
matter as a composite of such elementary `material' entities is essentially circular. 
 Similarly
the impression we have of matter on any scale as having an apparent independent %%@
existence or sense of inertia is only relative to other test or reference bodies, %%@
which are also assumed to possess similar innate `material' properties, and the %%@
hypothesis of independent material bodies on any scale cancels out by the circularity %%@
of the argument. The postulated material concept however remains of great pragmatic %%@
value in \textit{describing} the world and communicating information about it, as 
 for the empirical phenomena depicted in figure~\ref{arena}(c), while saying very %%@
little about what the physical world actually \textit{is} at a fundamental level.

  Indeed our prevailing understanding of the nature of matter as distributed in space %%@
has evolved significantly historically in time. The conception dating from Democritus %%@
(circa 460--370$\:${\small B.C.}) in ancient Greece proposing that everything is %%@
composed of indivisible and indestructible atoms of matter pursuing a pattern of %%@
motion according to deterministic natural laws 
  has remained influential. The laws of motion, based on quantitative empirical %%@
observations, were expressed with mathematical precision for the extended and %%@
impenetrable parts, subject to forces of attraction, composing all bodies in the %%@
Newtonian mechanical worldview of the $17^{\mathrm{th}}$~century.  
  Subsequently in the $19^{\mathrm{th}}$~century Faraday and Maxwell developed the %%@
field concept to account for electromagnetic phenomena, for which
 the notion of `action at a distance' between particles of matter through empty space %%@
could consequently be discarded. Maxwell's theory and equations for the %%@
electromagnetic field influenced Einstein's theory of the gravitational field, %%@
culminating in equation~\ref{Eineq} in the early $20^{\mathrm{th}}$~century. A unified %%@
field picture could then be sought with solutions for classical fields representing %%@
corpuscular states, either in terms of localised regions of high energy density in the %%@
electromagnetic field, as described by Weyl as noted above, or with massive particles %%@
corresponding to microscopic extreme structures of the gravitational field itself.

    Also in the first half of the $20^{\mathrm{th}}$~century a more ephemeral %%@
conception of matter was introduced with quantum mechanics and quantum field theory, %%@
with `particle' phenomena again ascribed to the properties of fields, now as
 `quanta' of field excitations, while in other
developments unification schemes with fields themselves deriving from the properties %%@
of extra spatial dimensions were first proposed by Kaluza and %%@
Klein~\cite{Kaluza,Klein}. In the latter half of the $20^{\mathrm{th}}$~century these %%@
two frameworks were combined in string theory, with the methods of quantum theory %%@
based upon a point-like particle model in 4-dimensional spacetime adapted and  applied %%@
consistently for one-dimensional relativistic vibrating `strings' in a 10 or %%@
26-dimensional spacetime, with particle states represented by quantised string %%@
excitations. As a leading candidate for a theory of `quantum gravity' the technical %%@
developments of string theory continue to progress in the $21^{\mathrm{st}}$~century, %%@
as we also discuss in the following section.

 In his historical review of western philosophy Bertrand Russell, in considering the %%@
meaning that might be attached to the word `matter', adopts a pragmatic approach in %%@
expressing the opinion: `My own definition of `matter' may seem unsatisfactory; I %%@
should define it as what satisfies the equations of physics' (\cite{Russ} book 3 %%@
towards the end of chapter 16). This raises the question regarding the purpose of %%@
theoretical physics itself, in terms of whether it concerns an ever evolving %%@
\textit{description} of matter, indefinitely refined by observations and an improving %%@
mathematical account, or whether the ultimate goal is to uncover an understanding of %%@
what matter actually \textit{is}, with material entities possessing a structure that %%@
can be considered isomorphically equivalent to the mathematical expressions of the %%@
theory.
 Indeed, while our theoretical conception of matter has evolved, we nevertheless tend %%@
to assume that there is a real objective sense in which `matter' in %%@
$5^{\mathrm{th}}$~century {\small B.C.} Athens is exactly the \textit{same} as that in  %%@
$21^{\mathrm{st}}$~century {\small A.D.} Princeton, having a coherent, rational %%@
essence, with only the domain of our knowledge having changed.

  For the theory proposed in this paper the ambition is to explain what matter %%@
actually \textit{is}, and \textit{why} it has the properties it is observed to have. %%@
This aim is based on the approach of developing the theory by beginning with an %%@
underlying conceptual motivation from which the mathematical structure of the theory %%@
and corresponding properties of matter are subsequently deduced, rather than by %%@
adopting a conception of matter that directly describes or parametrises observed %%@
empirical phenomena or by setting the theory within a largely internally motivated and %%@
sophisticated mathematical framework from its inception. This approach is summarised %%@
in the title of this paper, with the properties of matter together with the %%@
geometrical form of spacetime itself proposed to derive directly from the elementary 
 substructure of the underlying flow of time, as a universal unifying principle.

  The point of view adopted here regarding the relations between space, time and %%@
matter might then be considered as a non-trivial `gestalt shift' from a more standard %%@
conception of these structures. Here the emphasis is placed firmly on the pre-eminence %%@
of time, and our perception of it through the mathematical \textit{forms} of time, %%@
rather than focusing from the outset upon forms of matter \textit{in} space and time. %%@
This significant shift in perspective hinges on the interpretation of %%@
equation~\ref{propint}.

  A well known example of a gestalt shift involving our perception of an image is the %%@
drawing of the `duck-rabbit' as presented here in figure~\ref{dubbit}(b).

\begin{figure}[htbp]  
\centering
\epsfxsize=14.4cm
\leavevmode
\epsffile[0 0 2533 717]{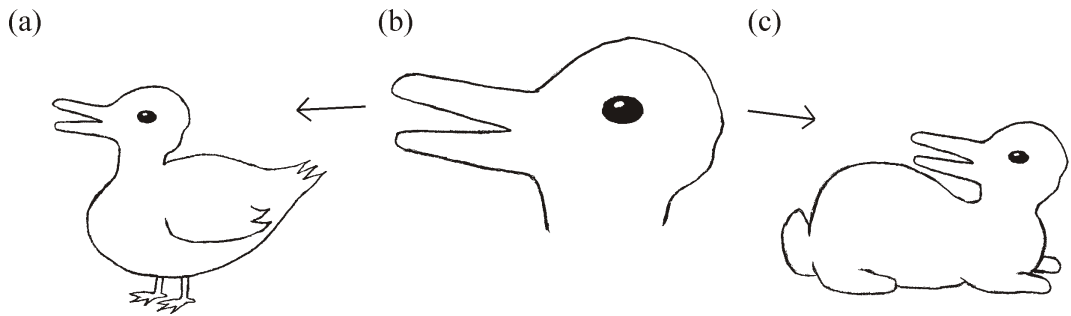}
\caption{\setb  (b) The central drawing can be seen either as the head of a duck %%@
facing to the left or of a rabbit facing to the right~(\protect\cite{drab}, 
 \protect\cite{Wittg} part II section XI), 
 with the mental change in perspective between these two possibilities known in %%@
psychology as a `gestalt shift' (from the German word `Gestalt' for `shape' or %%@
`form'). The corresponding extrapolations for the two interpretations are sketched in %%@
(a) and (c) respectively.}
\label{dubbit}
\end{figure}

  There is no objective fact of the matter regarding whether figure~\ref{dubbit}(b) %%@
depicts a duck or a rabbit, with the choice being purely a question of adopting a %%@
particular subjective perspective, which can be alternated. However once we have %%@
decided to see either a duck or a rabbit the details of an augmented drawing will %%@
diverge between the two cases as we extrapolate below the neckline of %%@
figure~\ref{dubbit}(b), with unambiguous differences between features of a duck or a %%@
rabbit emerging as depicted in figures~\ref{dubbit}(a) and \ref{dubbit}(c) %%@
respectively.

   In the case of equation~\ref{propint}, which encapsulates the structure of the %%@
local inertial frames implicit in figure~\ref{arena}(b), we can see this relation %%@
\textit{either} as an expression for `time' $\delta s$, by emphasising the left-hand %%@
side, or for `spacetime' $(\delta x^0, \delta x^1, \delta x^2, \delta x^3)$, by %%@
focusing on the right-hand side. If we adopt the latter perspective and extrapolate %%@
beyond equation~\ref{propint} for higher-dimensional spacetime structures the first %%@
step would be to add a single extra spatial dimension $x^4$ with a `$-(\delta x^4)^2$' %%@
term appended to the right-hand side as considered by Kaluza and Klein \cite{Kaluza, %%@
Klein}. With four of the additional components of the augmented 5-dimensional metric %%@
field interpreted as the electromagnetic 4-vector potential field
 $A_{\mu}(x)$, incorporated into a single framework alongside the original %%@
gravitational field components $g_{\mu\nu}(x)$, this initial step dating from the %%@
1920s was encouraging in terms of provisional connections with the empirical world as %%@
it was known then. Further augmentations in the structure of extra spatial dimensions %%@
have led to  a large range of possible models in recent decades, as alluded to in the %%@
opening of section~\ref{pers1}, for which however, as we also noted there,
 identifying a direct and unambiguous connection with specific properties of the %%@
modern-day Standard Model of particle physics has proved difficult.

  On the other hand on making a gestalt shift and looking at equation~\ref{propint} %%@
the other way, adopting the alternative interpretation of considering that equation to 
 represent a particular arithmetic expression for the form of time on the left-hand %%@
side, the extrapolation now leads to the generalisation of equation~\ref{propgen}. %%@
Rearranged as equation~\ref{lvo} this naturally leads to the explicit 56-dimensional %%@
quartic form of time in equation~\ref{lvfs}, as briefly reviewed in %%@
section~\ref{pers3}, with the breaking of the corresponding $\ese$ symmetry over a %%@
4-dimensional spacetime base exhibiting the structure of table~\ref{esebreak}. 
 Hence in exploring the natural mathematical extrapolation for this interpretation we %%@
directly identify features that closely resemble specific empirical properties of the %%@
Standard Model, including fractional charges and a left-right asymmetry.
 Beyond uncovering this rich vein of esoteric properties of the Standard Model,  we %%@
are led to a prediction of a yet higher-dimensional form of time with an $\ee$ %%@
symmetry to complete this empirical structure, as proposed in~(\cite{Unifi} section %%@
9.3) and explored further in~\cite{TimeE}.
  The achievement of these non-trivial inroads into the otherwise seemingly arbitrary %%@
and puzzling features of the Standard Model
  to a large degree vindicates this shift in perspective towards a unification scheme
    with both external spacetime and the observed properties of matter deriving from %%@
the arithmetic substructure of the underlying one-dimensional flow of time alone, and %%@
suggests that for establishing a fundamental physical theory this might ultimately be %%@
the right way to look at the world and to comprehend the underlying workings of the %%@
universe.

In terms of diagrammatic representation the most direct analogy with %%@
figure~\ref{dubbit}(b) is in the interpretation of figure~\ref{arena}(b). This latter %%@
diagram can be seen \textit{either} as a picture of the \textit{flow of time} $s$ %%@
manifested in the spacetime form $M_4$, \textit{or} as a picture of
\textit{spacetime} $M_4$ containing the flow of time $s$, with the choice of %%@
perspective hinging upon the interpretation of equation~\ref{propint} -- with the %%@
emphasis placed upon the left \textit{or} right-hand side respectively. This choice is %%@
augmented in figure~\ref{arena}(c) with either the \textit{flow of time} taking %%@
priority or with the emphasis placed upon \textit{spacetime and matter} collectively.  
 Here we adopt the former perspective, with the properties of both spacetime and %%@
matter carried within and deriving from the generalised form of time in %%@
equation~\ref{propgen}.

  Unlike the analogy of the duck-rabbit in figure~\ref{dubbit}(b), which involves %%@
different ways of looking at something \textit{in} the world, the shift in perspective %%@
described here for figure~\ref{arena}(c) regards the manner in which we see the %%@
\textit{whole} physical universe. This hence requires stepping back further from our %%@
preconceptions and assumptions made about the nature of the world itself. 
 In objective terms this gestalt shift is particularly hard to see since we %%@
immediately encounter physical structures in spacetime as apparently \textit{given}, %%@
out there for us to observe as depicted in figure~\ref{arena}(c). 
This situation is  somewhat 
 different to that regarding figure~\ref{dubbit}(b), for which the motivation for %%@
seeing  a duck or a rabbit is essentially a symmetric $50 \! : \! 50$ choice of %%@
perspective.

For the universe as a whole our tendency to see the world around us, as depicted for %%@
example in figure~\ref{arena}(c), as `matter in space and time' is a deep-seated, %%@
firmly rooted, viewpoint that we generally take for granted.
Inevitably the early developments in science set out by describing and cataloguing %%@
what we can physically detect, before seeking a deeper explanation for these empirical %%@
observations.
 However, from that perspective, beyond being posited for pragmatic purposes, it is %%@
difficult to conceive of what `matter' in space and time might fundamentally %%@
\textit{be} without forever begging the question of the nature of the next layer down,
 either literally or in an underlying explanatory sense. 
 Here we are suggesting a change in viewpoint from a basis of `matter in space and %%@
time' to a foundation in the `flow of time' alone. The corresponding 
 gestalt shift to this way of seeing the world as a manifestation of multi-dimensional %%@
forms of time remains however objectively somewhat counter-intuitive.

  It is from a subjective perspective, with all of our thoughts and observations of %%@
the physical world as represented in figure~\ref{arena}(c) flowing through and %%@
accompanied by an irreducible progression in time, as described near the opening of %%@
this section, that this change in viewpoint, with time playing the primary role, is %%@
perhaps more readily seen.
   Indeed, while we do not perceive extra spatial dimensions, and even find it very %%@
hard to conceive of what a higher-dimensional space would `look like', as also alluded %%@
to in the opening of this section,  we \textit{do} intimately experience the %%@
one-dimensional passage of time, and in this subjective sense this change in %%@
perspective is a natural one, and one that provides a simple and unambiguous starting %%@
point for a theory.
The corresponding mathematical basis for this theory is elementary, as seen in %%@
equations~\ref{propint}--\ref{lvo}, but is nevertheless accompanied by a significant %%@
and non-trivial gestalt shift towards this conceptually novel way of looking at the %%@
world.

 The approach could be taken of constructing a theory based on positing %%@
equation~\ref{lvo} as an `ansatz', then following the mathematical structures of %%@
equations~\ref{lvo}--\ref{lvfs} and exploring the consequences, without considering %%@
the underlying conceptual and philosophical elements of the motivation and %%@
interpretation for the theory. In this manner similar success in obtaining the %%@
symmetry breaking pattern of table~\ref{esebreak} and a foothold in the properties of %%@
the Standard Model could be achieved. However, this would miss the conceptual origin %%@
of the theory, which is seen as an essential and irreducible element of its %%@
foundation.  
 Alternatively, if we do start with this conceptually novel basis and set out to %%@
construct a physical theory purely out of the one-dimensional flow of time we are %%@
naturally led to these mathematical structures, which are found to exhibit this %%@
recognisable correspondence with empirical properties of particle physics.

  Speaking in the early 1920s Einstein~\cite{Einqu} noted that while the natural %%@
sciences gained a degree of security from applying mathematics, the connection between %%@
mathematics and the physical world remained an uncertain one.
  In general the observation of connections made between the mathematical structure of %%@
scientific theories and the physical structure of the empirical world has often been %%@
accompanied by an appreciative element of surprise, as famously pondered by 
 Einstein~\cite{Einqu}: 
\begin{quotation}
    How can it be that mathematics, being after all
a product of human thought which is independent of experience, is
so admirably appropriate to the objects of reality?
\end{quotation}

 Here we have adopted the perspective that an irreducible element of all `experience', %%@
and of all `human thought', is the passage of time. The one-dimensional continuum of %%@
time represented by $s \in \rrr$ provides the original link with a mathematical %%@
structure which by the \textit{defining} properties of the real line incorporates the %%@
elementary substructure of equation~\ref{propgen}, which can be written as %%@
equation~\ref{lvo}. The possible multi-dimensional forms of time in turn incorporate %%@
the basic geometric structure of \mbox{4-dimensional} spacetime, through the quadratic %%@
form in equation~\ref{lvfr}, together with natural extensions on to homogeneous cubic %%@
and  quartic expressions.    
 The symmetry breaking structure of explicit multi-dimensional forms of time, such as %%@
equation~\ref{lvfs}, necessarily projected over the substructure of the 4-dimensional %%@
spacetime arena, makes direct contact with elementary properties of the physical %%@
world, as deduced for example for table~\ref{esebreak}.

  That is, the substructure inherent in an interval of time $\delta s$ resembles the %%@
microscopic structure of the physical world as explored in high energy physics %%@
experiments. All properties of matter more generally are proposed to arise in this %%@
way, infused in, deriving from and intimately connected with the passage of time %%@
through which all physical objects of our experience are encountered, as represented %%@
in figure~\ref{arena}(c). Hence in principle this theory carries with it an %%@
explanation of how the mathematical structures deriving from it can  account for the %%@
structures of the empirical world in a less surprising manner through this intrinsic %%@
connection.

 Since the employment of the independent continuous variable of time in Newton's %%@
method of fluxions, which he invented and then applied to describe the motion of %%@
bodies through space in his mechanics, successful physical theories utilising %%@
differential calculus, including Maxwell's equations, the Dirac equation, quantum %%@
theory and general relativity, have incorporated the notion of a continuous flow of %%@
time as parametrised by a real number $s \in \rrr$. 
 The properties of this continuum are employed here in particular in deriving the %%@
general expression for the form of time in
  equation~\ref{lvo}, which relies on the infinitesimal nature of the real numbers. 
  This is similar to the way that other theories utilise the continuous progression of %%@
time in differential expressions, except that the present theory is founded upon the %%@
structure of the temporal continuum \textit{alone}. We conclude this section by %%@
summing up the essential argument for how such a construction is possible.

  When seen as an extension from the integers, with elements $p,q \in \zzz$, and the %%@
rational numbers, with elements expressed in the quotient form $\frac{p}{q} \in \qqq$
 for $q\neq 0$, the unique complete ordered field of the real number system $\rrr$,
  in containing the former cases as subsets, can be defined in turn through the %%@
relatively sophisticated, and isomorphically equivalent, constructions of Cantor, via %%@
Cauchy sequences of rational numbers, or 
 of Dedekind, via cuts partitioning the set of rational numbers (see for %%@
example~\cite{Wais} chapter 13, \cite{Smith} part I on Dedekind).
 However the intuitive notion of a gapless continuum is a very simple idea, as for the %%@
conception of progression in time.  While time itself is not a number,
the real numbers provide a rigorous mathematical representation of this %%@
one-dimensional ordered continuum as employed in physical theories. 

   This structure $\rrr$ may appear somewhat mysterious in comparison to the apparent %%@
substructures of $\zzz$ and $\qqq$, but these latter number systems are \textit{not} %%@
appropriate mathematical objects to represent the \textit{flow of time}, which is not %%@
here to be thought of as described by a `collection of points'. The notion of a `point %%@
in time' is mathematical idealisation or limiting extrapolation which does not embody %%@
the essential property of time which, as considered here, necessarily \textit{is} a %%@
one-dimensional continuum. The essence of time is  lost in extracting `a mathematical %%@
point' of time, as something that can never be subjectively encountered.  
  Here we begin with the concept of time as a one-dimensional gapless continuum and %%@
then utilise the real numbers as a structurally isomorphic mathematical model as the %%@
basis for a physical theory. The real numbers appear perhaps less mysterious when %%@
introduced through this conceptual motivation rather than in the mathematical context %%@
of other number systems. 
 The continuum property is the key feature, with
  actual `real numbers' only associated with arbitrary intervals of time
   for pragmatic purposes
    \textit{relative} to a particular unit, such as an Earth day, and only to within %%@
the limits of precision of measuring devices and to within the number of significant %%@
figures employed in the explicit decimal representation of the real number.

  When we think of the substructure of time we might first think of days containing %%@
hours containing minutes containing seconds and so on, according to conventional units %%@
for dividing up the real line representing time. However the structure of the real %%@
number system is much richer than this, in particular involving multiplicative as well %%@
as additive operations, containing the substructure of equation~\ref{propgen} at an %%@
elementary level for an infinitesimal real interval. Since the interval $\delta s \in %%@
\rrr$ \textit{has} this arithmetic property, and since
 this mathematical continuum is identified with
 the continuum of time, then time \textit{itself} can be considered to possess the %%@
richer substructure described in equation~\ref{propgen}. We can then ask how %%@
\textit{this} substructure of \textit{time} might be manifested.

  In particular since the possible arithmetic compositions of the interval $\delta s$ %%@
incorporate the quadratic metrical structure in equation~\ref{propint} and %%@
\ref{propeta} time itself carries with it a substructure that can be realised in a %%@
geometrical `spatial' form. This is the change in perspective we are adopting for %%@
figure~\ref{arena}(b).
 Through this basic arithmetic expression for a real interval the one-dimensional %%@
gapless continuum of time can be manifested as an indefinitely extended 4-dimensional %%@
gapless continuum of spacetime $M_4 \equiv \rrr^4$, deriving directly from the %%@
translation symmetry of equation~\ref{propint} and~\ref{propeta}, incorporating the %%@
geometric properties of 3-dimensional Euclidean space.

 Given the more general higher-dimensional and higher-order forms for time in %%@
equation~\ref{propgen} the \textit{necessity} of perceiving the world in %%@
\textit{space} as well as \textit{time} projects out the quadratic substructure which %%@
supports the spacetime base manifold $M_4$, maintaining a \textit{local} 4-dimensional %%@
pseudo-Euclidean structure,  as a framework for the observation of apparent forms of %%@
matter such as sketched in figure~\ref{arena}(c), with the properties of matter %%@
deriving from the residual temporal components and symmetry breaking pattern.  
   That is, by mapping the continuous flow of time, through which we perceive the %%@
physical world, in a structurally isomorphic one-to-one manner onto the real numbers a %%@
theoretical structure can be directly derived in purely mathematical terms that we can %%@
then \textit{map back} onto the empirical world in a structurally isomorphic %%@
one-to-one manner in principle in the form of a unified physical theory. The theory %%@
might then be tested against the empirical data to within the precision of calculation %%@
and experiment.

  The general form for the continuum of time in equation~\ref{propgen}  can be %%@
rewritten in terms of finite components in equation~\ref{lvo} and 
  explicit full forms of time with a high degree of symmetry considered, as described %%@
in section~\ref{pers3} for extensions from equation~\ref{lvfr} with
  Lorentz symmetry to equations~\ref{lvts} and \ref{lvfs} with an $\esi$ and $\ese$ %%@
symmetry respectively.
 Analysing the properties of the remnants surviving the $\ese$ symmetry breaking %%@
projection over $M_4$, as summarised in table~\ref{esebreak}, the elementary %%@
substructure of time is found to bear a close resemblance to the elementary %%@
microstructure of the physical world as observed in the high energy physics %%@
laboratory. In making this `gestalt shift' in perspective, and following through the %%@
consequences, it is striking that the development of this very simple idea, founded %%@
upon the one dimension of time alone, leads to a series of esoteric empirical %%@
properties of the Standard Model of particle physics, \textit{without} the need to %%@
postulate an independent material substratum to accommodate these properties.

  Time, from this perspective, is not just a benign independent parameter, not just a %%@
spectator of events passing in the world, but rather simultaneously underlies the %%@
geometric form and determines the empirical structure of all observations in the %%@
world. The temporal structure \textit{of} the world \textit{is} the world, and to %%@
adopt this perspective is to understand the basis for this theory.

%\pagebreak

\section{Establishing a Firm Foundation}
\label{pers5}

   Many unification frameworks, such as those that posit extra spatial dimensions as %%@
alluded to in the opening of section~\ref{pers1} \cite{Rizzo,Shif,Cheng,Liu}, set out %%@
by introducing extra structures over and above 4-dimensional spacetime.
 For some models with extra dimensions there are predictions for effects beyond the %%@
Standard Model that may be accessible to laboratory experiments, although no such %%@
effects have yet been empirically observed (see for example \cite{Hewett,Kret}). There %%@
is also perhaps a danger of overreaching without first making decisive progress in %%@
assimilating specific features of the Standard Model itself into the extra spatial %%@
dimensions paradigm, which has proved difficult as also noted in the opening of %%@
section~1 with reference to \cite{Witt,Jitt}.

  For the present theory, based in contrast on the single dimension of time, we have %%@
been able to utilise the rich structure of existing clues from observations in high %%@
energy physics experiments, as embodied in the Standard Model, as an empirical %%@
criterion to initially test the theory against. 
  The connections established at the level of the $\esi$ symmetry for the form of time %%@
in equation~\ref{lvts} include the identification of Weyl spinors and an internal %%@
$\suth_c \times \uo_Q$ symmetry with the appropriate fractional charge structure, as %%@
reviewed in (\cite{TimeE} subsection~4.2).
 The natural embedding of this cubic form in the quartic form of equation~\ref{lvfs} %%@
suggests that the latter can be interpreted as a higher-dimensional form of time with %%@
an $\ese$ symmetry, for which further connections with the Standard Model might be %%@
expected. This has been verified, with the further features of Dirac spinors and an %%@
intrinsic left-right asymmetry identified for this $\ese$ symmetry of time, as %%@
summarised here for table~\ref{esebreak} and reviewed in 
 (\cite{TimeE}~subsection~4.3). This connection with observations provides a %%@
reassuring foundation to build upon, leading to the theoretical prediction of a %%@
further augmentation to an $\ee$~symmetry of time as pursued in (\cite{TimeE} %%@
section~5), and in turn to potential empirical predictions as tentatively outlined in %%@
(\cite{TimeE} section~7).

 We also note that the present theory, based on multi-dimensional forms of time, is %%@
also very different from models with extra timelike dimensions (see for example %%@
\cite{Erdem,Quir} and the references therein). While maintaining a quadratic form
such models augment the 4-dimensional spacetime form on the right-hand side of %%@
equation~\ref{propeta} with a non-Lorentzian metric signature, with care then needed %%@
to avoid conflict with causality and unitarity. Here we are generalising the form of %%@
time itself, leading to cubic and higher-order polynomial forms as described for %%@
equation~\ref{propgen}. The identification of a smooth  4-dimensional spacetime %%@
background from a quadratic substructure, as the necessary arena for all observations, %%@
then breaks the full symmetry of time. 
  This 4-dimensional manifold $M_4$ incorporates a local metric with the Lorentz %%@
signature of equation~\ref{propeta} which determines a light cone structure on $M_4$ %%@
within which causal relations are well-defined, reflecting the underlying ordered %%@
one-dimensional progression in time itself (see also the discussion of causality %%@
in~\cite{Unifi} midway through section~13.3).

  This underlying motivation was contrasted with that of special relativity in %%@
section~\ref{pers2}. Drawing the two postulates of special relativity together in a %%@
consistent framework for electrodynamics the aesthetic guide of simplicity led %%@
Einstein to a clarification of the formulation of time and simultaneity (\cite{Einsr}, %%@
\cite{Pais} chapter~7(a)). With no absolute time or preferred reference frame defined %%@
in special relativity `there are as many times as there are inertial frames' %%@
(\cite{Pais} chapter~7(a)). This new approach to a physical theory was further %%@
developed in general relativity to incorporate in a consistent manner inertial frames %%@
that are not related by a uniform relative motion as well as arbitrary reference %%@
frames. For the present theory essentially `there are as many times as there are %%@
observers', each of whom is associated with a fundamental temporal parameter 
 $s \inn \rrr$ as represented in the centre of figure~\ref{arena}(c), and each of whom %%@
is carried inexorably into the future with the ordered causal progression in time %%@
encapsulated in its spacetime $M_4$ manifestation. The compatibility of this %%@
multiplicity of times, corresponding to a multiplicity of observers, and the mutual %%@
reciprocal relations between them in the full theory is very similar to that in %%@
special and general relativity (\cite{Unifi} ending of section~5.3 and near the %%@
opening of section~13.1).

  Extrapolating beyond the 4-dimensional form of time of equation~\ref{propeta}, and %%@
generalising beyond the local inertial frames of general relativity, we are led %%@
directly to equation~\ref{propgen} and observe that in projecting higher-dimensional %%@
forms of time over the \mbox{4-dimensional} spacetime manifold $M_4$ we obtain a %%@
`theory of matter' in spacetime, as described towards the end of section~\ref{pers3}. %%@
In this manner the structures arising directly from an observer's temporal flow $s$ %%@
include both matter fields and other forces of nature in addition to gravity, with the %%@
trajectory of the observer through spacetime 
  buffeted by the non-gravitational forces and  
 not in general pursuing the course of a local inertial reference frame -- a situation %%@
that arises in the vicinity of the Earth only under very special, or temporary, %%@
circumstances.

 The specific properties of matter derived will depend upon  the full form of time, %%@
that can be written as $\lvh$ as described for equation~\ref{lvo}, and the full %%@
symmetry $\hG$, which has been proposed to be the exceptional Lie group $\ee$ as noted %%@
above.
 In light of the above-mentioned issues of both causality and unitarity
we also note that  the particular non-compact real form of $\ee$ to be employed is
proposed to be obtained through augmentations from the 4-dimensional Lorentz group via %%@
the 10-dimensional Lorentz group, as described for (\cite{TimeE} equation~89), 
  with a compact internal symmetry group $G$ in equation~\ref{dirprod} required for a %%@
consistent quantum theory limit (\cite{TimeE} section~6). Even for the full quantum %%@
structure of this theory, deriving from equation~\ref{gfromavt} as described in %%@
section~\ref{pers3}, the external 4-dimensional spacetime $M_4$ with a local Lorentz %%@
symmetry is considered a smooth continuous base manifold structure. 
   
  In other theories spacetime itself may be composed of or exhibit an intrinsically %%@
discrete or grainy structure. This is the case for `loop quantum gravity' (see for %%@
example~\cite{Rove})  with `quanta of space', on a microscopic scale associated with %%@
the Planck length, represented by the nodes of a `spin network'. In this case the %%@
apparent features of a smooth spacetime only emerge on the macroscopic scale, which %%@
extends down to all scales presently observable. 
  The theory aims to construct a generalisation of quantum field theory
  without a background metric structure and
   consistent with general covariance, hence respecting this central symmetry of %%@
classical general relativity.
The philosophy adopted by loop quantum gravity is to tackle one major problem at a %%@
time, specifically the identification of a quantum field theory for which general %%@
relativity arises in the classical limit, while unification with the Standard Model is %%@
not incorporated within this picture. In this sense the aims are less ambitious than %%@
those of string theory (see for example \cite{Polch}), the other main candidate for a %%@
consistent quantisation of gravity. 

  Any proposed fundamental theory will ultimately 
 need to account for established successful theories, consistently combining
 general relativity with
  quantum field theory as applied for particle physics phenomena 
   together with an explanation of the Standard Model,
 at least in the appropriate limiting approximations consistent with all observations, 
 and ideally with some novel predictions empirically verified. 
   Even if a unification scheme should achieve these technical and empirical successes %%@
questions can still be raised concerning the \textit{origins} of the theory in terms %%@
of \textit{why} the world should be this way.
Given the focus upon questions at the other end, regarding for example the derivation %%@
of observed Standard Model properties, the foundational questions are sometimes %%@
postponed or overlooked. 
  This may leave a theory protractedly suspended upon the provisional basis of an %%@
ansatz or set of postulates that is declared this way since `we have to start %%@
somewhere',
 which seems insufficient for an `ultimate' theory.

        Questions regarding the ultimate origin of a theory, beyond its pragmatic %%@
utility, might in fact be considered intractable, prompting in some cases the %%@
subjective notion that the workings of the universe ought to be described by %%@
`aesthetically pleasing' mathematics, which might provide a guide towards constructing %%@
such a theory. This has also been the case in employing Lie groups for proposals of %%@
Standard Model unification, ranging from the early SU(5) `Grand Unified Theory' %%@
\cite{GeoGla} for which the authors propose from the outset that `the uniqueness and %%@
simplicity of our scheme are reasons enough that it be taken seriously', to the %%@
incorporation of gravity also in the $\ee$ model of \cite{Lisi} which opens with an %%@
appeal to the principle that `the mathematics of the universe should be beautiful'. %%@
However, while being of some heuristic value, such a criterion is neither well-defined %%@
nor decisive in pointing towards an ultimate unification scheme, and hence is not %%@
fully satisfactory
  in itself in motivating the basis for a theoretical framework.

  A good deal of work in theoretical physics involves addressing internal mathematical %%@
technicalities or problems that have arisen in developing the structure of existing %%@
theories, often with no immediate sight of either the foundational questions at the %%@
one end or connections with the empirical world at the other. This is perhaps the case %%@
for some of the progress made in developing string theory~\cite{Polch}, in pursuing %%@
the ambition of incorporating a consistent quantised theory of gravity. If this %%@
program is ultimately successful,
 even identifying one or more preferred string configurations that reproduce the %%@
properties of both the Standard Model of particle physics
 and  large scale cosmological structure out of a vast collection of possible %%@
solutions on addressing the `landscape problem'~\cite{Doug}, the question would %%@
\textit{still} remain concerning \textit{why} the world should be this way, apparently %%@
constructed from the fundamental objects of one-dimensional `strings' or %%@
higher-dimensional `branes' in a 10 or 11-dimensional spacetime for example.

 These foundational issues are perhaps exacerbated by the fact that historically %%@
string theory was discovered somewhat accidentally, having its roots in a different %%@
application as an unsuccessful model for hadrons from the 1960s, rather than a more %%@
direct motivation.
 All consistent string theories possess a closed string state describing a zero mass %%@
spin-2 particle, which is problematic for a model of hadrons and as such the theory %%@
was superseded by quantum chromodynamics. However in the 1970s it was `felt that %%@
string theory was too beautiful to be just a mathematical curiosity'~\cite{Schw}, and %%@
with the massless spin-2 state in principle describing the `graviton', the proposed %%@
carrier of the gravitational force, string theory was reinterpreted as a natural %%@
candidate for a fundamental theory of `quantum gravity' united with the quantum %%@
theories for the other forces of nature and matter fields.         
  In avoiding point-like particle entities string theory also brought with it
a softer short distance behaviour, in principle evading the calculational infinities %%@
that plagued other attempts to quantise gravity. 
  The conceptual motivation for string theory however still remains of a seemingly %%@
provisional nature into the $21^{\mathrm{st}}$ century, with the emphasis perhaps %%@
being placed more upon the rigor of the mathematical formulation of the theory, which %%@
is a somewhat novel approach compared to earlier developments in physics.

 In the case of general relativity by contrast a simple conceptual picture based on %%@
Einstein's insight into the intrinsic structure of spacetime as demonstrated by his %%@
`thought experiment' concerning the perspective of an observer in free fall came first %%@
in 1907; as described by Einstein as `the happiest thought of my life' (\cite{Pais} %%@
chapter 9)
 and encapsulated in the equivalence principle. This principle was itself motivated %%@
both by general experience and experimental observation of falling objects. There then %%@
followed several years of technical mathematical development in the geometric %%@
structure of the theory leading to the Einstein field equation~\ref{Eineq} and a %%@
theory of gravitation in 1915~(\cite{Eingr}, \cite{Pais} \mbox{chapters~9--14}).

 On the other hand the mathematical formulation of quantum theory was introduced in %%@
the mid-1920s based on innovation and a working set of assumptions,
 improvised by a number of physicists including the key figures of Heisenberg, 
  Schr\"{o}dinger, Born and Bohr, 
 driven by the empirically observed quantities it was designed to model  (see for %%@
example \cite{Bethe} chapter `Theory, Criticism and a Philosophy' by Heisenberg, %%@
\cite{Pais2}~chapter~12).
 Only after the mathematical scheme had been postulated and successful results %%@
achieved was the language developed to describe it, while the conceptual %%@
interpretation of quantum theory is still being debated today.
 The theory is nevertheless grounded in unequivocal laboratory observations.
 Heisenberg~\cite{Bethe} also explains his scepticism towards placing too much %%@
emphasis on rigorous mathematical methods, based on the concern of becoming too %%@
detached from the experimental data.

  This focus upon the mathematical scheme became increasing significant in developing %%@
the sophisticated calculational tools of quantum field theory (QFT) in the %%@
mid-$20^{\mathrm{th}}$ century,  which nevertheless have achieved considerable success %%@
in matching the measurements made in the high energy physics laboratory. 
 Without a firm conceptual underpinning and
with the technical formulations of theories seemingly taking on an independent life of %%@
their own it was in this context that Wigner wrote of the  `unreasonable effectiveness %%@
of mathematics in the natural sciences' \cite{Wign}.

  This sentiment echoes the `How can it be\ldots?' quote from Einstein cited in the %%@
previous section. It would seem all the more surprising that a mathematical theory %%@
should account for phenomena in the empirical world if both the founding motivation %%@
for the theory lacks a clear attachment to the physical world and the internal %%@
formalism of the theory has been developed in a similarly detached vein.
 While general relativity makes this connection with the physical world through the %%@
equivalence principle,
 quantum theory, in the form for example of Heisenberg's `matrix mechanics',  is %%@
rooted in the empirical observations it relates, in particular regarding patterns of %%@
atomic spectral data. In both cases 
 significant empirical successes have been achieved beyond the original scope of the %%@
theories and without meeting any failures.

 On the other hand formulations of quantum gravity, such as string theory or loop %%@
quantum gravity, arguably lack either a conceptual or observational anchor in the %%@
physical world, being founded largely upon addressing the technical challenges arising %%@
from the assumption that gravity should be quantised, and empirical successes for %%@
these theories have to date been limited. While possessed of elegant and sophisticated %%@
mathematical structure, some of which does reflect our knowledge of the physical %%@
world, it may be that the development of these frameworks, and even of QFT itself, %%@
despite the technical and pragmatic successes, may have been somewhat premature in %%@
lacking the support of a firm conceptual basis.

  It is sometimes suggested that a final unifying theory is still very remote from us %%@
by a considerable amount of further work and technical breakthroughs into the future %%@
(\cite{Hooft} see for example the contribution from Rovelli), or even that the goal of %%@
a single unifying theory may be untenable~\cite{Hawk}.
  These views are typically expressed with reference to the status, rate of progress, %%@
and presently perceived obstacles in the context of an existing theoretical framework, %%@
such as string theory or loop quantum gravity, which may indeed be some distance from %%@
providing an ultimate resolution. In principle however a new idea offering a new %%@
perspective has the potential of providing a different path towards that same shared %%@
ultimate goal, along which the obstacles may not appear so insurmountable, bringing %%@
the prospects of a complete unified theory much closer than otherwise anticipated.

 In particular, with respect to foundational questions, compared with string theory %%@
the situation is essentially diametrically opposite for the theory described in this %%@
paper. Here from the beginning we consider the conceptual and philosophical questions %%@
concerning what a theory might look like in order to explain \textit{why} the universe %%@
should be this way. Posed in the context of studying theories based on extra spatial %%@
dimensions we make a subtle change in perspective in founding the theory upon the flow %%@
of time alone, as the elementary one-dimensional continuum through which all of our %%@
observations are made.
Pursuing the elementary mathematical expression of this idea 
 the implicit substructure of an interval of time 
 can provide the source of both spacetime and the matter it contains, via the %%@
structure and interpretation of equations~\ref{propint}--\ref{lvo}.
 Explicit mathematical forms have \textit{then} been identified and applied to fill %%@
out this conceptual picture, rather than moulding  the development of the theory from %%@
the outset within the confines of a preconceived or postulated mathematical framework.
 With mathematics providing a precise extension of familiar spoken language in order %%@
for a theoretical framework to connect with and describe the physical world,
 and to understand what the theory \textit{means},
 it should ideally be built upon the support of a rational underlying conceptual %%@
picture -- one that can be comprehended and conveyed in unambiguous linguistic terms %%@
and which itself exhibits a manifest connection with the world. 
 \mbox{Based on} the firm conceptual foundation of a single dimension of time the %%@
present theory is also consistent with the view expressed by Einstein~\cite{Einqu2}:
\begin{quotation}
  It can scarcely be denied that the supreme goal of all theory is to make the 
irreducible basic elements as simple and as few as possible without having to
 surrender the adequate representation of a single datum of \mbox{experience}.
\end{quotation}

  This sentiment is often paraphrased as the maxim: `Everything should be made as %%@
simple as possible, but not simpler'. As for the quote from Einstein discussed here in %%@
the previous section the above quote is frequently cited by theoretical physicists in %%@
the $21^{\mathrm{st}}$ century -- they are included in this paper to reflect %%@
inclinations common in modern-day theoretical physics as much as those of Einstein %%@
nearly a hundred years ago. 

  The foundation of the present theory can be compared with the origins of general %%@
relativity, developed from a largely conceptual basis, and that of quantum theory, %%@
motivated mainly from an empirical basis, as reviewed earlier in this section.
 Here the original motivation is not based upon a particular kind of experience or %%@
upon particular experiments but rather we simply note that \textit{all} experience and %%@
\textit{all} experiments take place irreducibly \textit{in} time and that time %%@
contains \textit{within} itself a substructure that can be expressed mathematically %%@
and utilised to construct a full physical theory of the world.     
  In addition to the underlying conceptual and mathematical simplicity
 elements of mathematical naturalness and uniqueness are in part employed in leading %%@
from equation~\ref{lvo} via equation~\ref{lvfr} to equations~\ref{lvts} and %%@
\ref{lvfs}, with the actions of the respective Lie groups $\esi$ and $\ese$ describing %%@
a high degree of symmetry for these multi-dimensional forms of time. In this manner %%@
contact is made with both familiar structures from the mathematical physics literature %%@
as well as with empirical structures of the physical world as summarised in %%@
table~\ref{esebreak}.

  The present theory is hence firmly grounded at \textit{both} ends, being %%@
conceptually founded on the simple notion of the flow of time through which all %%@
observations are made through to the successes achieved in accounting for a series of %%@
empirical features of the Standard Model of particle physics (with further possible %%@
empirical connections reviewed in section~\ref{pers3} and cited in %%@
section~\ref{pers6}).
 There is also a close and transparent connection between these two ends, with %%@
properties of the Standard Model deriving \textit{directly} from the symmetry breaking %%@
pattern for the full form of time of equation~\ref{lvfs}, which is motivated as a %%@
natural instantiation for the general form of time of equation~\ref{lvo} which, as the %%@
central equation of this theory, provides a \textit{direct} mathematical expression of %%@
the underlying conceptual picture. 
 This firm foundation in both the conceptual \textit{and} the empirical sense, %%@
\textit{together} with the close relation between them, then provides
 a robust basis for the further mathematical and technical development of the theory.

 The explanatory power of the theory leads also to predictive power in pointing to a %%@
role for $\ee$ as the ultimate symmetry of time~\cite{TimeE} as recalled in the %%@
opening of this section. As reviewed in 
 (\cite{TimeE} subsection~2.3) real forms of the exceptional Lie groups $\esi$, $\ese$ %%@
and $\ee$ are known to describe symmetries associated with certain natural %%@
mathematical generalisations of 4-dimensional spacetime
 (see for example~\cite{Gunay2}). For $\esi$ and $\ese$ these structures can also be %%@
interpreted as symmetries of time, for the forms of equations~\ref{lvts} and %%@
\ref{lvfs} respectively, which naturally incorporate as a substructure  4-dimensional %%@
spacetime and the Lorentz symmetry of equation~\ref{lvfr}.
 These observations in part motivate considering the largest and unique exceptional %%@
Lie group $\ee$ as the symmetry of the full homogeneous polynomial form of time, %%@
accommodating the subgroup chain $\ee \supset \ese \supset \esi \supset %%@
\mbox{Lorentz}$, in each case with a symmetry breaking pattern deriving from the %%@
necessary projection over the external 4-dimensional spacetime substructure with the  %%@
Lorentz symmetry subgroup acting on the local tangent space $\TM_4$ of the spacetime %%@
manifold $M_4$. 
 While analysis of the $\esi$ and $\ese$ stages has already in principle provided %%@
explanations  for several puzzling features of the Standard Model, the breaking of the %%@
predicted full $\ee$ symmetry group is proposed to complete the full Standard Model %%@
particle multiplet picture, as we argue in~\cite{TimeE}.
 The identification of the precise structure of this $\ee$ action and the specific %%@
composition of the full form $\lvh$ itself, with the appropriate properties, then %%@
remains as a theoretical puzzle to be addressed.

  As well as the natural mathematical embeddings in the progression towards %%@
higher-dimensional forms of time and the unique mathematical structures involved, we %%@
might also in principle attempt to associate these structures with a notion of %%@
`mathematical beauty', if such a concept might be correlated with possession of a high %%@
degree of symmetry. There are four classical Lie groups which, as for the largest %%@
exceptional Lie group $\ee$, are associated with a Lie algebra having an 8-dimensional %%@
maximal Abelian subalgebra, which is hence a common core feature of these five groups. %%@
These four rank-8 classical Lie algebras are A$_8$ (su(9)), B$_8$ (so(17)), C$_8$ %%@
(sp(16)) and \mbox{D$_8$ (so(16))}, composed respectively of a total of 80, 136, 136 %%@
and 120 independent symmetry generators. By comparison the rank-8 exceptional Lie %%@
algebra $\ee$ comprises a total of 248 generators, and hence describes a greater  %%@
concentration of independent symmetry actions that could be interpreted as quantifying %%@
a higher degree of mathematical beauty.
 We note, however, that while such an aesthetically appealing  property is perhaps %%@
desirable it is not our primary guiding motivation here.

   This progression towards a significant role for $\ee$ as a symmetry of time can be %%@
considered then as a testable \textit{theoretical} prediction of the theory (as noted %%@
in~\cite{TimeE} at the end of subsection~5.1). This in turn is sufficient to give a %%@
hint towards the potential \textit{empirical} predictions for the theory (as listed %%@
in~\cite{TimeE} section~7).
   As well as the predicted application of an $\ee$ symmetry the precise nature of %%@
`quantisation' for the theory is currently the other main area of focus in developing %%@
this framework; we have touched upon both of these connections with physics here in %%@
section~\ref{pers3}.

   As noted there a picture has emerged in which the gravitational field itself is %%@
\textit{not} quantised. With quantum theory based on a set of postulates we don't %%@
actually know \textit{why} 
 anything is subject to quantum theory, so the assumption that everything, including %%@
the gravitational field, should be quantised seems highly provisional. On the other %%@
hand since the gravitational field can be identified with the 4-dimensional spacetime %%@
geometry, and since all matter is in spacetime, in this sense everything \textit{is} %%@
covered by gravity.
Rather than incorporating gravity under an all-embracing set of postulates of quantum %%@
theory, the present  theory can be considered more as a generalisation of general %%@
relativity (as described in~\cite{TimeE} opening of section~3 and alluded to here %%@
before equation~\ref{propeta}).
 Here gravitation provides the source and explanation of the quantisation of all %%@
non-gravitational fields, through the local degeneracy in underlying field solutions %%@
for identifying  the geometry of the external spacetime $M_4$ itself,  as described %%@
for equation~\ref{gfromavt} in section~\ref{pers3}, with the standard machinery of QFT %%@
proposed to arise in the flat spacetime limit.

 From this point of view the motivation for constructing a consistent theory %%@
incorporating `quantum gravity', and 
  the technical difficulties that arise from the assumption that gravity should be %%@
quantised, are no longer a concern. 
 This then marks another significant difference with the origins of string theory, for %%@
which the consistent quantisation of gravity is a central goal. 
  The existence of `gravitons' would of course be very difficult to empirically %%@
verify, other than perhaps through the proposal that such hypothetical quantum %%@
fluctuations in the gravitational field in the very early universe might be greatly %%@
amplified by an inflationary phase and observed today via signals of a classical %%@
cosmological gravity wave background~\cite{Wilc}.
     Theories of quantum gravity that imply a discrete or `foamy' texture for %%@
spacetime itself on the Planck scale, such as loop quantum gravity, might also be %%@
experimentally probed~\cite{Chou}.
 Advances in technology have already in principle 
  brought each of these potential signatures for quantised gravity or spacetime within %%@
reach,
  with constraints being placed in the absence of 
  any clear signals~\cite{Bicep,Chou2}.
 Hence the picture that has emerged for this aspect of the present theory in which the %%@
gravitational field is not quantised and spacetime is considered smooth down to %%@
arbitrary scales is also testable, in the non-trivial sense of being in principle %%@
`falsifiable' given the potential for observations to the contrary. In the meantime, %%@
while the data remains inconclusive,
 it is perhaps in any case worth exploring theoretical frameworks both for which %%@
gravity is and is not quantised.

  For the present theory, with standard QFT being an effective theory arising in the %%@
flat spacetime limit, it is also the case that quantum phenomena for non-gravitational %%@
fields in a highly curved spacetime will need to be understood, and may well differ %%@
from the predictions that have been obtained from formulations of QFT applied in such %%@
an acutely non-flat background environment. Indeed there are theoretical issues that %%@
remain to be resolved in modelling quantum particle phenomena for an extreme spacetime %%@
geometry (see for example~\cite{Mathur}). For the new framework we  will need to  %%@
reassess fundamental questions such as how, and even whether,  black holes radiate and %%@
lose mass, and the issues raised such as the `information paradox'.  The need for a %%@
coherent description of extreme gravitational regions, such as in the vicinity of the %%@
`singularity' of a black hole or the Big Bang, for which classical general relativity %%@
alone is ultimately insufficient, and  to correspondingly incorporate quantum %%@
phenomena in a theoretically consistent manner into this picture, then in principle %%@
provides a further ambitious theoretical test for this framework.

  On the other hand, despite the differences, there is a significant overlap between %%@
the mathematical structures we have been led to for the present theory and elements of %%@
the mathematical formalism employed in frameworks such as string theory (as noted for %%@
example in~\cite{TimeE} sections~2 and 6). While there remains a degree of debate over %%@
the merits of string theory as an ultimate unified theory (see for %%@
example~\cite{Woit,Duff}) diverse applications in physics and mathematics associated %%@
with a repurposing of string theory or M-theory have been identified.
  More generally, much of the mathematics literature that we have employed, including %%@
elements of  \cite{Man4,Wang2,Borst,Rios,Gunay2} as alluded to here in the opening of %%@
section~\ref{pers3} and above, has been in part motivated by various developments in %%@
theoretical physics  in recent years.  
 Closer examination of the nature of these mathematical connections might in principle %%@
prove to be mutually beneficial in contributing to the understanding and development %%@
of both the present and other theoretical frameworks, in particular with the common %%@
goal of a complete unified theory in mind.

 For the present theory while the simplicity of the basic underlying idea, expressed %%@
in the form of equation~\ref{lvo},  and non-trivial structural correspondence  %%@
identified with the Standard Model,
 via equation~\ref{lvfs} and table~\ref{esebreak},
 provide a robust basis, other areas of the theory for which progress has been made, %%@
including the relation between gravity and quantum phenomena
  centring  upon equation~\ref{gfromavt}, 
 are mathematically at a provisional stage requiring further development. In the %%@
meantime, while borrowing related mathematical structures from other theories, a %%@
significant contribution from the present framework is in identifying a means of %%@
establishing a firm and unambiguous conceptual foundation with a direct link to the %%@
empirical successes. This has been achieved through a change in perspective in placing %%@
the flow of time, and its possible multi-dimensional manifestations, at the heart of %%@
the theory.

%\pagebreak

\vspace{-1pt}
\section{Conclusions}
\label{pers6}

  In concluding we place this paper, the central theme of which has been the nature of %%@
the `gestalt shift' in perspective from matter in spacetime towards a fundamental role %%@
for time, in the context of our earlier papers that have developed this theory. 

  The technical details underlying the connections made between the theory and the %%@
Standard Model of particle physics, as outlined here for table~\ref{esebreak}, are %%@
described in (\cite{Unifi}~chapters~6--9) as summarised in \cite{Novel} and reviewed %%@
in \cite{TimeE}, with further analysis and emphasis upon the predicted role for $\ee$ %%@
in the latter reference. The question of the ultimate symmetry and specific structure %%@
of the corresponding full multi-dimensional form  for time is an open one. In the %%@
previous section of this paper we have considered the high degree of symmetry %%@
exhibited by the exceptional Lie groups as a possible factor, and while other factors %%@
are described in the above papers a more complete understanding is desired.  The %%@
plausibility of identifying an underlying explanation for `quantisation' through %%@
equation~\ref{gfromavt} is explored in detail in (\cite{Unifi} chapters 10~and~11), %%@
through a close comparison with the canonical formalism of QFT, and further elaborated %%@
in (\cite{KKone}~subsection~5.3,
  \cite{TimeE} section~6).

 Progress has also been made in incorporating some of the existing geometrical %%@
techniques employed for Kaluza-Klein theories, as developed in (\cite{Unifi} chapters %%@
2--5) and analysed further and more succinctly in \cite{KKone}, as alluded to here %%@
before equation~\ref{gfromavt}. Possible contributions to questions concerning the %%@
large scale structure of the universe, including potential `dark sector' candidates %%@
deriving from the bottom line of table~\ref{esebreak}, are described in (\cite{Unifi} %%@
chapters 12 and 13) in the context of the standard model of cosmology, as summarised %%@
in (\cite{TimeE} towards the end of section~6).

   Overall the change in perspective emphasised in this paper has proved very %%@
fruitful, with the length of \cite{Unifi} in part reflecting the author's attempt to %%@
examine a wide range of the low-hanging fruit within reasonable reach. The prediction %%@
of the $\ee$ symmetry of time further developed in \cite{TimeE} points towards an %%@
ambition of grasping one of the higher branches of the theory, as does the proposal of %%@
providing a consistent unified framework for gravity and quantum theory. Progress may %%@
be needed in both of these areas in the upper canopy of 
 the theory in order to more fully reproduce high energy physics phenomena and make %%@
decisive predictions (see for example \cite{Unifi} section~15.2).
 In this paper we have returned to the roots of the theory in expanding upon its %%@
robust and firm foundations.

 Based simply upon the substructure of the one dimension of time alone 
 the most explicit success to date for this theory has been in the uncovering of %%@
several distinctive features of the Standard Model of particle physics,
  seen to  emerge in a more direct and transparent manner than for models based upon %%@
the introduction of extra spatial dimensions. More generally,
all branches of the present theory covered in \cite{Unifi,Novel,KKone,TimeE} are %%@
directly relevant to the aim of accounting for structures of the physical world %%@
unfolding from the one-dimensional flow of time, with all areas under development and %%@
with open questions remaining as we have attempted to describe in the papers. However, %%@
the observation that such a simple theory, based on such a simple idea, can have %%@
something to say about all of these corners of the empirical world on all scales is %%@
noteworthy for this proposed unification scheme.
The adoption of this change in perspective on the universe, in placing time at the %%@
foundation of the theory, is then further justified by this
 broad range of applicability and potential for further advances.

%\pagebreak

%\vspace{-30pt}
%\input{persbib.tex}
%Bibliography

% fit references shorter space
{\small
%\vspace{-1pt}
%{\setlength{\baselineskip}{0.94\baselineskip}
{\setlength{\baselineskip}{0.58\baselineskip}

\par}
}
%Bibliography

%\pagebreak

%%
\par}% \linespread{1.0} for main text (28/11/15)
%match '{\setlength{\baselineskip}{0.625\baselineskip}' above

\end{document}